\documentclass{IEEEoj}
\usepackage{latexsym}
\usepackage{graphicx}
\usepackage{amsfonts,amssymb,amsmath}
\usepackage{mathtools, cuted}
\usepackage{hyperref}
\usepackage{amsmath}
\usepackage[T1]{fontenc}
\usepackage{cite}
\usepackage{subcaption}
\usepackage{comment}
\usepackage{amsthm}
\usepackage{multirow}
\usepackage{overpic}
\usepackage{steinmetz}
\usepackage{array}
\usepackage{url}
\usepackage{color}
\usepackage{tikz}

\usepackage{algorithm}
\usepackage{algorithmic}
\usepackage{units}
\usepackage{mathrsfs}

\theoremstyle{plain}

\usepackage{textcomp}
\def\BibTeX{{\rm B\kern-.05em{\sc i\kern-.025em b}\kern-.08em
    T\kern-.1667em\lower.7ex\hbox{E}\kern-.125emX}}
\AtBeginDocument{\definecolor{ojcolor}{cmyk}{0.93,0.59,0.15,0.02}}
\def\OJlogo{\vspace{-4pt}\hskip-4pt\includegraphics[height=18pt]{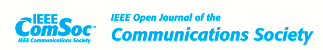}}
\begin{document}
\receiveddate{10 March, 2025}
\reviseddate{12 May, 2025}
\accepteddate{28 May, 2025}
\publisheddate{XX Month, XXXX}
\currentdate{3 June, 2025}
\doiinfo{OJCOMS.2025.3576931}
\title{Enabling 6G Performance in the\\ Upper Mid-Band by Transitioning\\ From Massive to Gigantic MIMO}

\author{Emil Bj{\"o}rnson\IEEEauthorrefmark{1} \IEEEmembership{(Fellow, IEEE)}, Ferdi Kara\IEEEauthorrefmark{1,2} \IEEEmembership{(Senior Member, IEEE)}, Nikolaos Kolomvakis\IEEEauthorrefmark{1} \IEEEmembership{(Member, IEEE)}, Alva Kosasih\IEEEauthorrefmark{1,3} \IEEEmembership{(Member, IEEE)}, Parisa Ramezani\IEEEauthorrefmark{1} \IEEEmembership{(Member, IEEE)}, and Murat Babek Salman\IEEEauthorrefmark{1}\IEEEmembership{(Member, IEEE)}}

\affil{KTH Royal Institute of Technology, Sweden.}
\affil{Zonguldak B\"ulent Ecevit University, T\"urkiye.}
\affil{Nokia Standards, Espoo, Finland.}
\corresp{CORRESPONDING AUTHOR: Emil Bj{\"o}rnson (e-mail:emilbjo@kth.se).)}
\authornote{This work is supported by the FFL18-0277 and SUCCESS projects from the Swedish Foundation for Strategic Research, the Grant 2022-04222 from the Swedish Research Council, and the SweWIN Vinnova Competence Center.}

\begin{abstract}
The initial 6G networks will likely operate in the upper mid-band (7-24 GHz), which has decent propagation conditions but underwhelming new spectrum availability.
In this paper, we explore whether we can anyway reach the ambitious 6G performance goals by evolving the multiple-input multiple-output (MIMO) technology from massive in 5G to gigantic in 6G.
We describe how many antennas are needed to reach the envisioned 6G peak user rates, how many can realistically be deployed in practical radio equipment, and what the practical spatial degrees-of-freedom might become. We further suggest a new deployment strategy that enables the utilization of radiative near-field effects in these bands for precise beamfocusing, localization, and sensing from a single base station site.
Finally, we identify open research and standardization challenges that must be overcome to efficiently use gigantic MIMO dimensions in 6G from hardware, cost, and algorithmic perspectives. 
\end{abstract}

\maketitle

\section{INTRODUCTION}
Massive multiple-input multiple-output (mMIMO) technology has been fundamental to fifth-generation (5G) wireless networks, which use it for beamforming, interference suppression, and spatial multiplexing \cite{2019_Björnson_DSP}. mMIMO has enabled higher user data rates and increased network capacity by greatly improving spectral efficiency compared to fourth-generation (4G) networks. The typical mMIMO configurations have 16, 32, or 64 antenna branches per base station (BS), and the average capacity gains compared to 4G is $20\times$, toward which mMIMO and broader bandwidths contribute almost equally \cite{Nokia_white_paper}.
The commercialization of mMIMO in 5G has been successful: At the end of 2023, mobile network operators (MNOs) provided mMIMO 5G coverage to 45\% of the world's population; in fact, the coverage is above 85\% in regions like North America, India, and China \cite{Ericsson2024}.

Present 5G deployments mainly use lower mid-band frequencies in the 3.5 GHz range and continue to expand worldwide. The next step is to use the 6.5 GHz band, which was recently identified for worldwide use by the International Telecommunications Union (ITU) \cite{WRC23}. Meanwhile, research into the next generation of wireless communication systems, known as 6G, is shifting towards standardization and technology development.
Evidently, 6G must provide more capacity to meet the ever-increasing traffic demands and enable new data-intensive applications such as virtual/augmented reality (VR/AR) and ultra-high-definition video streaming. To achieve commercial viability, 6G must also support new use cases and functionalities and ensure connectivity for many more devices with diverse characteristics.

Previous generational shifts have been associated with wider bandwidths at higher carrier frequencies because more spectrum directly leads to higher rates.
In the last decade, it was believed that the millimeter-wave (mmWave) band (24 GHz and above) would be the natural next step due to the abundant spectrum availability.
Hence, 5G was co-designed for two frequency ranges (FRs): FR1 with 0.4-7 GHz and FR2 with 24-71 GHz. However,  intermittent coverage caused by high susceptibility to blockage in mmWave bands presents a significant challenge, requiring expensive dense small-cell deployments to address. In light of these issues, the MNOs in South Korea closed their commercial mmWave networks in 2023.
Although the research community is actively investigating even higher bands, called terahertz (THz) and sub-THz, these will inherit and worsen the challenges and limitations of mmWave frequencies.

\subsection{Motivation and Related Works}
Towards this background, there is a growing interest from academia and industry in the missing interval along the frequency axis: the \emph{upper mid-band} from 7-24 GHz (also known as FR3) \cite{Nokia_uppermidband}. 
There are several key factors behind this development. Firstly, these frequencies offer significantly larger bandwidth than in the 3.5 GHz band. Moreover, the shorter wavelength enables MIMO configurations with many more antennas than in 5G but in a compact form factor. This allows for highly focused beamforming, enables spatial multiplexing, and generally reduces interference. Additionally, upper mid-band frequencies have more favorable propagation characteristics than mmWave frequencies, providing substantially better coverage and less sensitivity to beam misalignment \cite{Nokia_uppermidband}. 

The FR3 band is now known as the \textit{Golden Band} for 6G due to its promising balance between bandwidth availability and coverage \cite{cui20236g}.
\begin{figure*}
\centering
\begin{overpic}[width=0.95\textwidth]{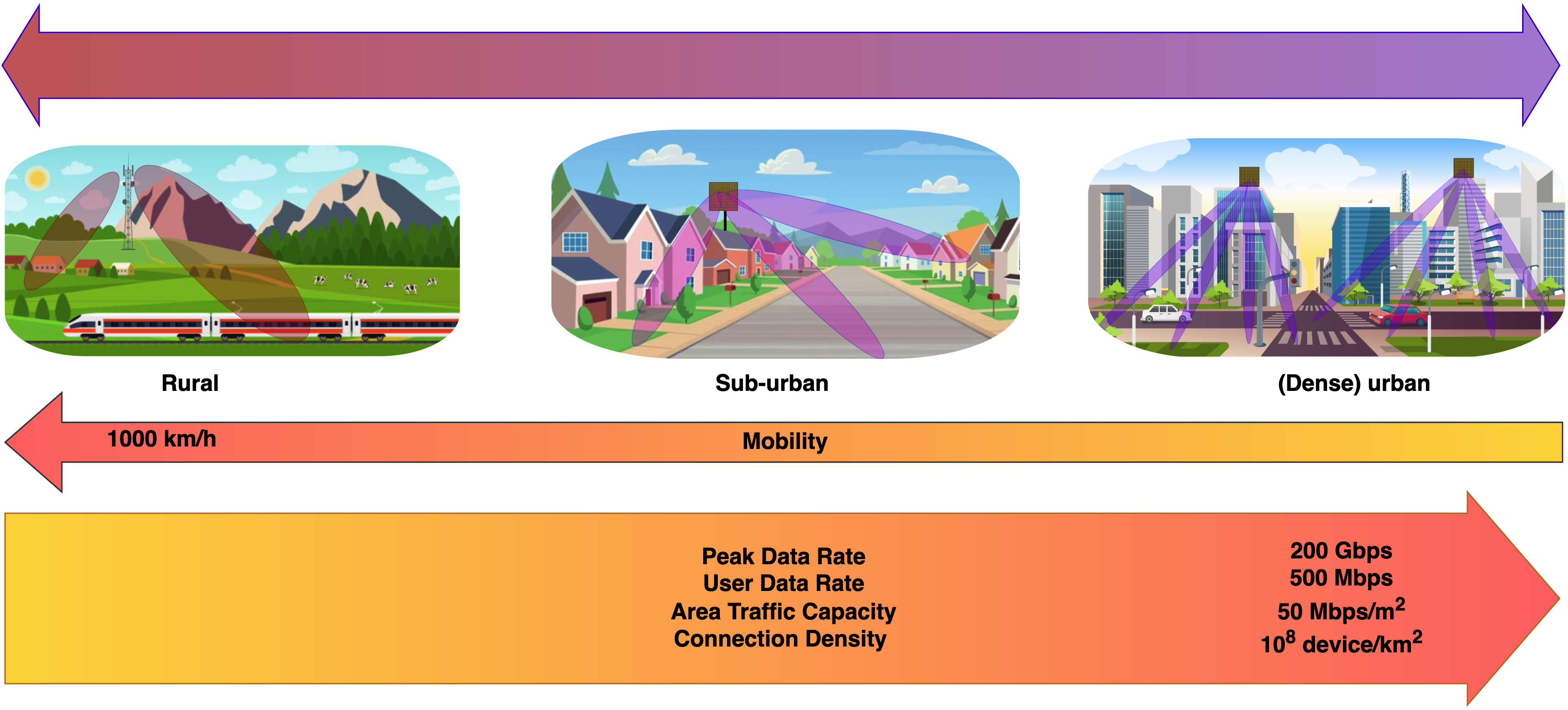}%
\put(4,47.5){\color{red}\textbf{3.5 GHz}}%
\put(6,38.5){\linethickness{.5mm}\color{red}\vector(0,1){8.3}}
\put(78,47.5){\color{red}\textbf{15 GHz}}%
\put(80,43.5){\linethickness{.3mm}\color{blue}\vector(1,0){7}}
\put(80,43.5){\linethickness{.3mm}\color{blue}\vector(-1,0){7}}
\put(70,44.5){\color{blue} 14.8 GHz}%
\put(83,44.5){\color{blue} 15.35 GHz}%
\put(80,38.5){\linethickness{.3mm}\color{red}\vector(0,1){8.3}}
\put(48,47.5){\color{red}\textbf{7.8 GHz}}%
\put(50,38.5){\linethickness{.5mm}\color{red}\vector(0,1){8.3}}
\put(50,43.5){\linethickness{.3mm}\color{black}\vector(1,0){7}}
\put(50,43.5){\linethickness{.3mm}\color{black}\vector(-1,0){7}}
\put(38,44.5){\color{black} 7.125 GHz}%
\put(53,44.5){\color{black} 8.4 GHz}%
\put(20,43.5){\linethickness{.3mm}\color{red}\vector(1,0){7}}
\put(20,43.5){\linethickness{.3mm}\color{red}\vector(-1,0){7}}
\put(10,44.5){\color{red} 4.4 GHz}%
\put(23,44.5){\color{red} 4.8 GHz}%
\end{overpic}
\caption{The new potential 6G frequency bands considered by ITU and their relation to IMT 2030  performance targets. }
\label{fig:system_model}
\end{figure*}
The 6G standardization begins in 2025 and will finish in 2028-2029, so commercial 6G networks can open in 2029-2030. These networks will likely operate in the upper mid-band. In this context, many researchers have started to study this band, including achievable communication performance and applications where it can be utilized. Not only the wireless communications community \cite{Kang2024cellular,cui20236g,new_mid_band_magazine,bazzi2025uppermidbandspectrum6g,lópezpérez2024capacitypowerconsumptionmultilayer,kang2024spectrumsharing,testolina2024,lee2022, shakya2024wideband,hu2024channelmodelingfr3upper,shakya2024propagationmeasurementschannelmodels,bomfin2024experimentalmultibandchannelcharacterization,shakya2024urbanoutdoorpropagationmeasurements,abbasi2024ultrawidebanddoubledirectionallyresolvedchannel,shakya2024angularspreadstatistics675,rasteh2024nearfieldmeasurementuppermidband}, but also the RF/microwave community \cite{kim2024,ghotbi2023,wong2023a,wong2023b} has put great effort into developing advanced antennas and filters operating in the bands, since the production of 6G systems will be completed soon. In \cite{Kang2024cellular,cui20236g,new_mid_band_magazine,bazzi2025uppermidbandspectrum6g,lópezpérez2024capacitypowerconsumptionmultilayer}, the authors discuss the conceptualization of upper mid-band 6G systems along with their visions. In \cite{kang2024spectrumsharing,testolina2024}, the co-existence of upper mid-band 6G systems with existing networks (i.e., satellite) is discussed, and some spectrum sharing mechanisms are proposed.  On the other hand, most of the articles in wireless communications for the upper mid-band are devoted to identifying channel characteristics in \cite{lee2022, shakya2024wideband,hu2024channelmodelingfr3upper,shakya2024propagationmeasurementschannelmodels,bomfin2024experimentalmultibandchannelcharacterization,shakya2024urbanoutdoorpropagationmeasurements,abbasi2024ultrawidebanddoubledirectionallyresolvedchannel,shakya2024angularspreadstatistics675,rasteh2024nearfieldmeasurementuppermidband}. These studies are commonly based on measurements or ray tracing to identify some critical characteristics of this band such as path loss, shadowing, angular spread, near-field effects, etc. Moreover, in \cite{raviv2024localization,mezzavila2024sensing}, the authors explore the sensing capabilities in the upper mid-band. As seen in the discussion of the literature, although channel characteristics, co-existence with existing services, and coverage assessment in this new band were recently reviewed, the MIMO assessments in this band have not yet been explored. To the best of the authors' knowledge, the only paper in the literature that deals with MIMO in the upper mid-band is \cite{heath2024beamsharing}, where the authors proposed a user-pairing and using a shared beam for near-field and far-field users in a MIMO setup. Therefore, realizing the full potential of mMIMO in the upper mid-band frequencies for 6G requires further studies.

\subsection{Contributions}

This tutorial paper describes why major improvements in MIMO technology are the key to delivering the promised 6G performance enhancements. {Although mMIMO originally refers to systems with an unlimited number of antennas \cite{marzetta}, the term is now commercially used to describe MIMO configurations with 16, 32, or 64 antenna ports in 5G. The recent research literature contains alternative terms associated with MIMO in THz, continuous apertures, or extremely large physical sizes---none of which is the topic of this paper or likely to be considered in the early days of 6G. Therefore, we believe that a new term is needed to refer to commercial 6G MIMO configurations in the upper-mid band. We propose the term \emph{gigantic MIMO (gMIMO)}, indicating an even larger number of antennas than in mMIMO. We advocate for this term to be used in systems with at least 256 antenna ports.}

In this paper, we explore the potential of gMIMO in the upper mid-band spectrum and examine the communication and localization performance in this frequency range. We start by identifying the candidate frequencies and available bandwidth within the upper mid-band spectrum to deploy gMIMO and outline the attractive features of this frequency range to meet related key performance indicators in ITU recommendation \cite{ITU2023}. Next, we analyze the number of antennas required in an array operating at upper mid-band frequencies to maintain the same path loss as experienced in the mid-band frequency range and assess the directionality of a typical upper mid-band array by examining its half-power beamwidth (HPBW). We then focus on the communication performance of upper mid-band frequencies in a gMIMO setup and evaluate whether this new spectrum can meet 6G achievable rate requirements. Additionally, we investigate upper mid-band near-field effects and explore the potential for beamfocusing and near-field localization and sensing in this Golden Band. Here, we demonstrate how distributed antenna arrays on the same site can achieve near-field communications and sensing performances. 
{Prior works on upper mid-band systems mainly provide overviews on communications, channel characterizations, spectrum co-existence/sharing in these bands. By contrast, this paper is the first to conduct an in-depth examination of the practical deployment of MIMO in the upper mid-band, with dimensions far exceeding current commercial setups. Our study uncovers the largely untapped potential of gMIMO in this new band and highlights the key challenges that must be addressed to realize its full promise in 6G. By bridging theoretical insights with practical deployment considerations, this paper aims to shape future design paradigms and drive the technological innovations needed for next-generation wireless systems.}

\subsection{Organization}
The remainder of this paper is as follows. In Section II, we review prospective 6G bands and their characteristics. In Section III, we elaborate on how many antennas we need to use in the upper mid-band, depending on what the design goal is. In Section IV, we further describe how gMIMO can enable enormous peak data rates and discuss the practically achievable performance. We then shift focus to the radiative near-field and how its related characteristics can be utilized to enable beamfocusing and precise localization in practical 6G deployments. This is covered in Section V and Section VI, respectively. In Section VII, we identify unexplored research challenges whose solutions could greatly impact 6G standardization and implementation. Finally, Section VIII concludes the paper.

\section{CANDIDATE 6G BANDS AND THEIR FEATURES}
\label{S_Ferdi}

The ITU harmonizes worldwide spectrum utilization, which is vital for efficiently deploying new wireless technologies, ensuring compatibility and interoperability across various regions. Hence, the ITU plays a crucial role in identifying new frequency bands for 6G. 
ITU calls cellular network technology \emph{International Mobile Telecommunications (IMT)} and has published its IMT-2030 vision for 6G \cite{ITU2023}. In this vision document, the ITU also points out some expectations and recommendations for 6G deployments. Table \ref{ITU} gives the most essential performance capabilities.
This section will briefly describe the specific 6G bands proposed by the ITU-organized World Radiocommunication Conference 2023 (WRC-23) and their properties compared to the 5G bands.

Taking into account factors such as anticipated technological advancements and the vision of IMT-2030, regional conflicts over existing wireless services, and demands from industrial stakeholders, WRC-23 identified three candidate bands (one from FR1 and two from FR3) for 6G, in addition to the continued use of legacy 5G bands \cite{WRC23}. These bands are as follows:\footnote{ITU Region 1 includes Europe, the Middle East and Africa, Region 2 includes the Americas, and Region 3 includes East Asia and Oceania.}

\begin{itemize}
    \item
\textit{4.4-4.8 GHz (Regions 1 and 3):} This additional sub-7 GHz band was requested by industrial stakeholders to provide extensive coverage and support high mobility in suburban and rural areas. 

    \item
\textit{7.75-8.4 GHz (Region 1) and 7.125-8.4 MHz (Regions 2 and 3):} This upper mid-band frequency is intended for outdoor-indoor applications in urban and suburban areas to meet high data rate and modest mobility requirements.

    \item
\textit{14.8-15.35 GHz (All ITU regions):} This upper mid-band allocation is mainly for dense urban and urban areas to meet the needs for massive connectivity and high data rates under relatively low user mobility.

\end{itemize}

The per-band use cases are illustrated in Fig.~\ref{fig:system_model}, which indicates how lower bands can manage higher mobility and higher bands support higher rates, capacity, and connection density. 
Importantly, not all IMT-2030 performance targets must be reached in all bands. These newly allocated upper mid-band frequencies offer several benefits for 6G deployments. In the following. we categorize how the upper mid-band makes differences from the existing sub-6 GHz or (anticipated and failed) mmWave bands.

\textbf{Co-existence and Interference Management:} %
Before any candidate bands can be assigned for 6G, further technical studies must be performed to determine if the intended performance can be achieved while ensuring co-existence with existing services, such as broadcast satellites and feeder links for satellite communication. These issues are detailed in \cite{cui20236g,Kang2024cellular, new_mid_band_magazine}. 
Traditional co-existence solutions are based on spectrum masks, antenna tilting, and cognitive radio. 
It is generally easier to enable co-existence between fixed transceiver infrastructures than in mobile communications, where the user devices can be anywhere and rotate arbitrarily.
Nevertheless, thanks to narrow and directive beams, the upper mid-band frequencies offer better isolation between adjacent channels, reducing interference, and making it easier to manage co-existence with other wireless systems. This also creates a perfect synergy between the upper-mid band frequencies and MIMO applications to utilize a spectrum- and energy-efficient wireless architecture.
Next-generation MIMO technology could offer a more flexible solution with extremely narrow directional beams and real-time out-of-system interference suppression.
ITU aims to finalize the technical studies by the next WRC in 2027, the last WRC before the intended 6G rollout in 2030.

\textbf{Spectrum Availability and its Consequences:}
The growing demand for data-intensive applications, such as VR/AR and ultra-high-definition video streaming, requires high data rates. This can be achieved by either extremely high spectral efficiency or broader bandwidths. {A common belief is that with a higher carrier frequency, wider contiguous bandwidths will be available, enabling higher achievable data rates. However, the three new bands only contain 400 MHz, 650-1275 MHz, and 550 MHz bandwidths, respectively. 
Depending on the region, the total spectrum is 1600-2225 MHz and will likely shrink substantially to resolve co-existence issues with incumbent systems already operating in these bands.
These numbers should be compared to the 500 MHz of the current 5G mid-band spectrum in the range 3.3-3.8 GHz (the exact numbers vary between regions), plus the new 700 MHz spectrum in the range 6.4-7.1 GHz assigned for 5G at WRC-23.
Hence, the first 6G BSs using one of the mentioned new bands will offer roughly as much new spectrum as the 5G networks in 2029 already have aggregated.
This shows that increased spectrum availability will not be the main distinguishing factor between 6G and 5G.
The ITU vision for 6G is a peak user data rate of 200 Gbps, a guaranteed user-experienced rate of 500 Mbps, and an area traffic capacity of 50 Mb/s/m$^2$ \cite{ITU2023}. Suppose 1.2 GHz of new spectrum becomes available, then a peak spectral efficiency of 166 b/s/Hz is required to reach the envisioned peak rate. This can only be achieved through major advances in MIMO technology. We will return to this matter in Section~\ref{S_Babek}.}

By enabling higher data rates than in 5G, the 6G network also obtains the ability to provide lower latency and/or higher reliability. The latency is reduced by transmitting shorter data blocks, which require more coding redundancy to maintain the same decoding reliability (i.e., block error rate). Alternatively, higher reliability is achieved by using stronger channel codes with more redundancy.

\textbf{Propagation and Hardware Characteristics:}
The upper mid-band frequencies offer a balance between capacity and coverage. As demonstrated later in this paper, we can reach substantially higher capacity than in current sub-7 GHz networks while maintaining better coverage than in mmWave bands. 
Signals in the upper mid-band can penetrate buildings more effectively than mmWave signals and experience propagation losses only slightly worse than in the FR1 bands \cite{shakya2024wideband,Nokia_uppermidband}. 
Hence, it might be possible to maintain the same BS site grid as in 5G.
This is key to making large-scale 6G deployments commercially viable, as the operational costs and coverage could be similar to 5G.

Another important characteristic is that MIMO technology in the upper mid-band can be implemented using fully digital beamforming. This is a major benefit over the power-hungry and inflexible hybrid beamforming solutions that are currently used in mmWave bands and contributed to the underwhelming practical performance of mmWave networks.

\begin{table}[]
\centering
\begin{tabular}{|ll|}
\hline
\multicolumn{1}{|l|}{Peak data rate per user (Gbit/s)}                             & 50--200               \\ \hline
\multicolumn{1}{|l|}{User experienced data rate (Mbit/s)}                             & 300--500                  \\ \hline
\multicolumn{1}{|l|}{Area traffic capacity (Mbit/s/m$^2$)} & 30, 50                               \\ \hline
\multicolumn{1}{|l|}{Connection density (devices/km$^2$)}  & $10^6$--$10^8$     \\ \hline
\multicolumn{1}{|l|}{Mobility (km/h)}                                     & 500--1000                 \\ \hline
\multicolumn{1}{|l|}{Latency (ms)}                                        & 0.1--1                         \\ \hline
\multicolumn{1}{|l|}{Reliability (error rate)}                            & $10^{-5}$--$10^{-7}$ \\ \hline
\multicolumn{1}{|l|}{Positioning accuracy (cm)}                                    & 1--10                                                  \\ \hline
\end{tabular}
\caption{The ITU Recommendations for IMT-2030 performance capabilities \cite{ITU2023}, including a range of target values for different deployment scenarios.}
\label{ITU}
\end{table}

\section{HOW MANY ANTENNAS DO WE WANT IN 6G BASE STATIONS AND DEVICES?}
\label{S_Nikos}

{It is often claimed that increasing the carrier frequency leads to higher pathlosses}, but the reality is more complex because pathloss depends on both the transceiver
hardware and free-space propagation. Understanding this trade-off is essential for designing 6G systems that operate at higher frequencies than 5G while achieving comparable coverage.

Suppose the transmitter and receiver have antenna arrays with effective areas $A_t$ and $A_r$, respectively.
In a free-space line-of-sight (LOS) scenario, the received power is $P_t A_rA_t / (d\lambda)^2$, which depends on the transmit power $P_t$, propagation distance $d$, and wavelength $\lambda$ \cite{massivemimobook}.
If the areas $A_t$ and $A_r$ are fixed, then the received power increases quadratically with the carrier frequency because $\lambda$ shrinks accordingly. Hence, it is better to operate at higher frequencies than lower ones. 
{In practice, each antenna element has an effective area proportional to $\lambda^2$, so the total effective area $A_t$ scales as $N_t\,\lambda^2$, which depends on the number of antenna elements, $N_t$, in the array.
Consequently, to keep $A_t$ (and similarly $A_r$) fixed as $\lambda$ shrinks, each array’s number of elements must increase proportionally to ${1}/{\lambda^2}.$}
For example, if the wavelength is halved (doubling the frequency), the effective area of each antenna element is reduced by a factor of four. To compensate, we need four times as many antennas to obtain the same total area. This would happen if we switch from the $3.5$ to the $7$ GHz band.

{This observation contradicts the common claim that ``increasing the carrier frequency leads to higher pathlosses'' mentioned at the beginning of this section.
To reach this opposite conclusion, one must consider a bad system design (e.g., with a fixed number of antennas, leading to smaller array areas as the frequency increases)} or non-LOS scenarios with materials that cause propagation losses that grow rapidly with frequency. The latter is the main practical showstopper for mmWave networks because mmWave signals are almost entirely blocked by concrete and experience an average indoor-to-outdoor loss that is more than 10\,dB higher than in FR1 \cite{Nokia_uppermidband}.

\begin{figure} 
    \centering
    \includegraphics[width=0.48\textwidth]{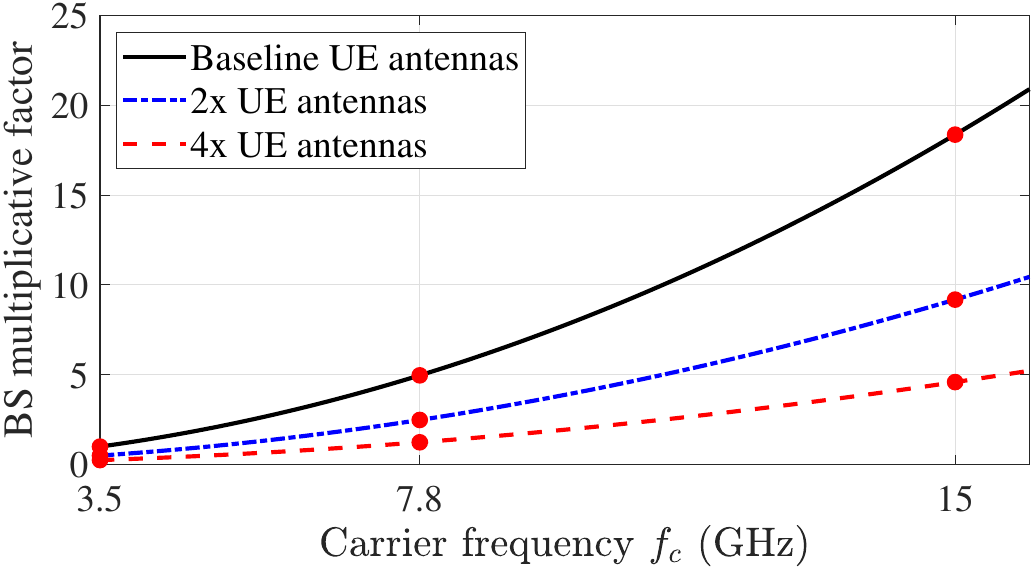} \vspace{1mm}
    \caption{The multiplicative factor for the number of BS antennas versus the carrier frequency. It expresses how many times more BS antennas are needed to maintain the same free-space pathloss as at $3.5$ GHz. The value depends on whether we scale the number of UE antennas.}
    \label{fig:antennas-pathloss}
\end{figure}

A potential design goal is to keep the free-space pathloss constant as the frequency increases, allowing us to reduce the array areas such that $A_tA_r\sim \lambda^2$. 
For instance, we can let the BS maintain its array area while the user equipment (UE) keeps its antenna number fixed.
Fig.~\ref{fig:antennas-pathloss} shows the multiplicative factor for the number of BS antennas that achieve the same pathloss at a carrier frequency $f_c$ as in a baseline case of $f_0=3.5$ GHz. 
There are curves for the cases with a fixed number of user-equipment (UE) antennas and $2\times$ or $4\times$ more antennas than in the baseline.
A BS in the 7.8 GHz band or 15 GHz band can have up to 5 or 18 times more antennas, respectively, than in current 5G networks.
If the UE doubles or quadruples its antenna elements, these numbers reduce to $1.2\times$--$2.5 \times$ and $4.6\times$--$9.2\times$, respectively.

\begin{figure}
\centering \vspace{4mm}
\begin{overpic}[width=\columnwidth]{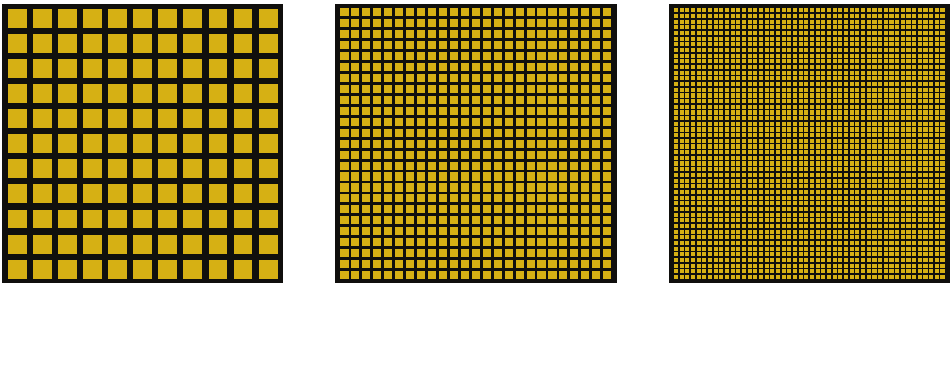}%
\put(2.5,42){mMIMO today}%
\put(43,42){gMIMO}%
\put(78.5,42){gMIMO}%
\put(7,5){$3.5$ GHz:}%
\put(42,5){$7.8$ GHz:}%
\put(78,5){$15$ GHz:}%
\put(3.5,0){$11 \times 11$ array}%
\put(39,0){$25 \times 25$ array}%
\put(74,0){$48 \times 48$ array}%
\end{overpic} \vspace{1mm}
\caption{{Comparison of three arrays at different carrier frequencies, but each having the same physical dimension of $0.5\, {\mathrm{m}} \times 0.5\, {\mathrm{m}}$}. The number of half-wavelength-spaced antennas that fit into this aperture grows quadratically with the frequency.}
\label{fig:array_magazine_paper}
\end{figure}

Even if the number of antenna elements grows quadratically with the carrier frequency, the physical size will not increase. Fig.~\ref{fig:array_magazine_paper} illustrates this fact for an array of dimensions $0.5\, {\mathrm{m}} \times 0.5\, {\mathrm{m}}$, which resembles  the current 5G BSs \cite{2019_Björnson_DSP}. With a typical half-wavelength spacing, this area fits $121$ dual-polarized antenna elements at 3.5 GHz, $625$ elements at 7.8 GHz, and $2304$ elements at 15 GHz. The number of antenna ports (i.e., transceiver branches) might be around 33\% of this because the industry usually connects each port to a group of 2-4 elements to limit the number of hardware components. {Regardless, the MIMO dimensions will increase when shifting from the 3.5 GHz band to the upper mid-band.}

{While the original mMIMO concept \cite{marzetta} assumed an arbitrarily large number of antennas to exploit favorable propagation and spatial multiplexing, practical deployments in FR1 (and FR2) are currently limited to 64 antenna ports due to hardware and reference signaling constraints. From the commercial viewpoint, the MIMO and mMIMO terms are intimately linked to 4G and 5G systems, respectively. The former is mainly used to provide beamforming gains, while the latter also enables spatial multiplexing in commercial products.   To distinguish future systems that push beyond this contemporary ceiling, especially in upper mid-band frequencies, we adopt the term gMIMO to describe the new MIMO configurations that we believe will be used in 6G systems. 
Since the currently largest mMIMO configuration at 3.5 GHz has 64 antenna ports and the upper mid-band enables at least $2^2=4$ times more ports in the same form factor, we suggest the new gMIMO terminology is used for systems with at least 256 antenna ports.
Importantly, the distinction is not only in numbers, but gMIMO will also feature new technological challenges and performance expectations described in the remainder of this paper.}

The pathloss discussion above only captures one MIMO functionality: beamforming. 
If we shift focus to spatial multiplexing, the rank of the channel matrix between the BS and UE---called the spatial degrees-of-freedom (DOF)---is limited by the number of UE antennas.
Even if not needed from a pathloss perspective, we can add more UE antennas to achieve a higher DOF, which typically leads to higher user rates in multipath environments.
On the other hand, from a channel estimation perspective, having many antennas only on one side (i.e., the BS) is advantageous because the estimation overhead is proportional to the minimum number of antennas at the transmitter and receiver \cite{massivemimobook}.
Hence, there is a non-trivial trade-off and
the preferred solution is scenario-dependent: the richness of the multipath environment determines how many large singular values can be achieved, and user mobility determines how many resources can be dedicated to channel estimation.

Increasing the number of antenna elements also impacts the beamwidth of the signals transmitted and received by the array. The half-power beamwidth, which is the angular interval where the radiation pattern is at least 50\% of its peak value, becomes narrower as the number of antenna elements increases. If we keep the array dimensions constant, the beamwidth is proportional to the wavelength, so that it will be $2.2\times$ smaller at 7.8 GHz and $4.3\times$ smaller at 15 GHz compared to the 3.5 GHz band.
The increased directionality reduces interference to/from unwanted directions but increases the need for precise and timely channel estimation.
This is cumbersome for both digital transceivers that cannot rely on beamforming gains during channel estimation and hybrid transceivers that require sophisticated tracking and beam-steering technologies to maintain robust communication links under mobility.

Finally, there are two good reasons for not keeping the free-space pathloss constant when moving to the upper mid-band but improving it by maintaining the array areas on both sides. Firstly, we might want a pathloss margin to handle larger non-LOS losses. Secondly, the signal-to-noise ratio (SNR) is inversely proportional to the bandwidth. If we cannot afford to increase the transmit power in 6G systems---even if we have more bandwidth---the extra beamforming gain comes in handy. This is particularly important in the uplink since the maximum power per UE is regulated.

\section{ACHIEVABLE RATES IN PROSPECTIVE 6G BANDS}
\label{S_Babek}

The user rates that can be achieved in a specific band vary enormously between the theoretical peak values and the average or guaranteed values under practical conditions. In this section, we will elaborate on both aspects for the prospective 6G bands in FR3.

The theoretical peak rate is computed as the product of the bit rate per symbol, the spatial DOF, and the bandwidth. Starting with the former, 5G supports up to 1024-QAM, while WiFi 7 also handles 4096-QAM. Assuming that 6G follows this development, we can reach 12 bit/symbol. Moreover, a current 5G device has 4 antennas, which could increase by $4\times$, leading to 16 spatial DOF.
Finally, if a single operator uses all 1.2 GHz of spectrum in the 7.8 GHz band, we will reach $12 \cdot 16 \cdot 1.2 = 230$ Gbps. Hence, the ITU vision of 200 Gbps from Table~\ref{ITU} is within theoretical reach but only by combining enormous spatial multiplexing with greatly expanded bandwidths.
The peak rate is reduced by one-third if the available bandwidth is divided between three MNOs. However, gMIMO gains remain feasible and essential to make the 6G user rates significantly higher than in 5G, and theoretical peak rates are only meant to be achievable in extreme test scenarios.

The upper mid-band is claimed to be a golden opportunity for maintaining the favorable coverage characteristics of sub-7 GHz bands while delivering typical user rates resembling or surpassing those in mmWave bands \cite{cui20236g}. To evaluate this claim, we have simulated a realistic propagation scenario. We can surely fit more antennas into a BS and device in FR3 than in FR1 to obtain higher beamforming gains, but there will be higher penetration losses. The question is: will the propagation be sufficiently favorable to obtain a higher data rate per UE, when taking all characteristics into account?

\begin{figure}
    \centering
    \subfloat [CDF of the data rate]
{\includegraphics[width=0.48\textwidth]{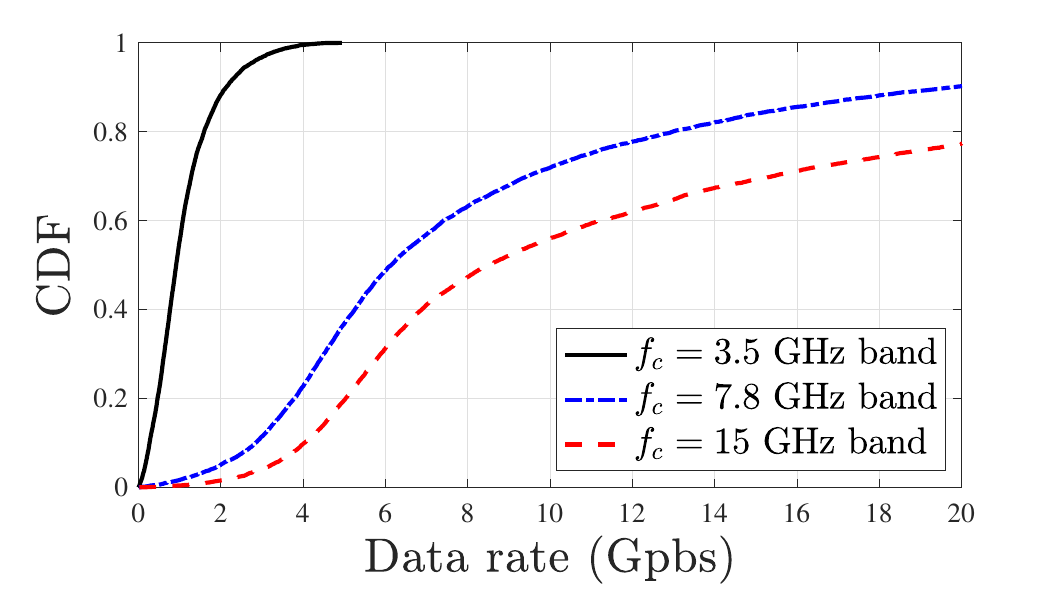}}\hfill
    \centering
    \subfloat[CDF of the number of transmitted streams]
{\includegraphics[width=0.48\textwidth]{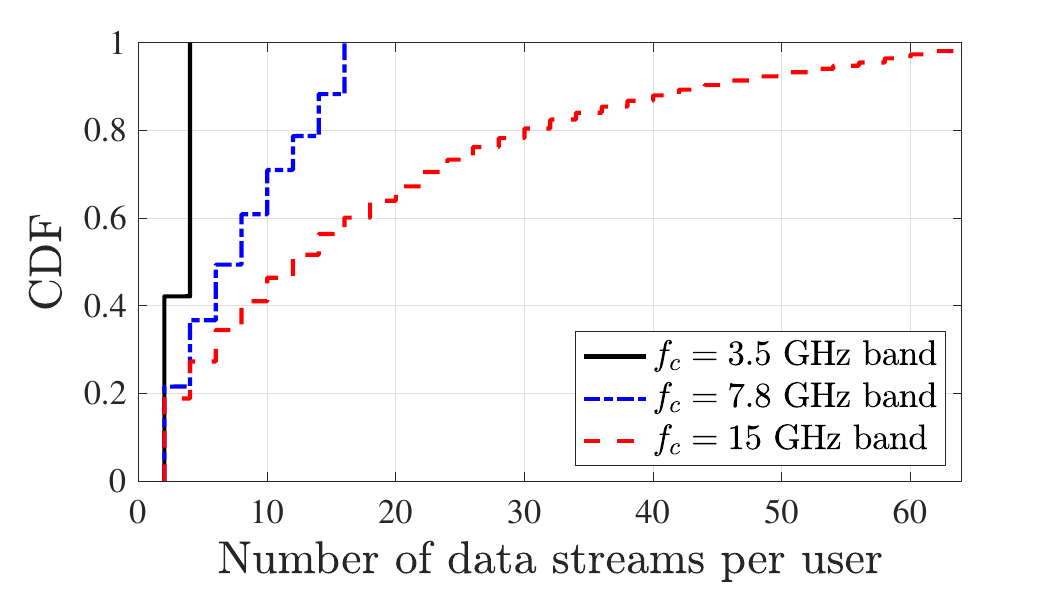}
     }
    \caption{CDFs of the per-user data rate and the assigned per-user data streams for spatial multiplexing. We considered a downlink LOS 3GPP Urban Micro Cell scenario. }
    \label{CapacityLos}
\end{figure}

We will compare the downlink user rates achievable in the new 7.8 and 15 GHz bands, shown in Fig.~\ref{fig:system_model}, with the current 3.5 GHz bands.
The size of the BS array is fixed to $0.5\, {\mathrm{m}} \times 0.5\, {\mathrm{m}}$; hence, the number of half-wavelength-spaced dual-polarized antennas increases with the carrier frequency. We consider $10$ UEs, each equipped with an antenna array of size $4\, {\mathrm{cm}} \times 8\, {\mathrm{cm}}$. The QuaDRiGa model developed at Fraunhofer HHI was used to generate realistic channel realizations for an urban microcell scenario \cite{Quadriga}.
The users are randomly distributed in a $1 \, {\rm km} \times 1 \, {\rm km}$ area around the BS. Fig.~\ref{CapacityLos}~(a) presents the cumulative distribution function (CDF) of the user rates obtained at different locations. We consider LOS scenarios and assume 400 MHz user bandwidth for upper mid-band frequencies and 100 MHz for sub-6 GHz frequencies. The transmit power of the BS is adjusted such that 1\,W is supplied per 1\,MHz bandwidth. Perfect channel state information (CSI) is assumed in the BS, and block diagonalization precoding based on water-filling power allocation is performed. 

The simulation results presented in Fig.~\ref{CapacityLos} demonstrate that the utilization of new frequency bands can significantly enhance the capacity for LOS scenarios compared to the sub-7 GHz bands. 1\,Gbps is guaranteed for almost all UEs, and many achieve more than 10\,Gbps.
This improvement is primarily due to the increased beamforming gain, which is facilitated by the ability to deploy a larger number of antennas in these new frequency bands. {The performance aligns with the anticipated requirements for future 6G systems, as discussed in Section \ref{S_Ferdi}, which aim to guarantee 500\,Mbps to everyone}. Due to the limited number of antennas that can be accommodated in the 3.5 GHz band, its capability to mitigate multi-user interference is constrained. Therefore, the new upper mid-band frequency shows promise in achieving improved capacity performance thanks to its ability to support more simultaneous user streams for LOS channels. Furthermore, the number of data streams that can be transmitted to each UE is investigated in Fig.~\ref{CapacityLos}~(b). $2$ streams per user ($1$ for each polarization) is seen to be utilized by most UEs at 3.5\,GHz. However, as the carrier frequency increases, more data streams can be provided per UE, which explains the shape of the curves in Fig.~\ref{CapacityLos}~(b). For the $7.8$ GHz band, up to $16$ data streams, and for the $15$ GHz band, up to $64$ data streams can be supported. These results demonstrate that higher DOF is achieved in practice when increasing the number of UE antennas, even if the array area is constant. While there is a considerable enhancement in the potential spatial multiplexing for individual users, the capacity improvement resulting from approximately doubling the center frequency from $3.5$ GHz to $7.8$ GHz is more significant than a comparable relative increase from $7.8$ GHz to $15$ GHz, primarily due to pathloss and available bandwidth. Consequently, the candidate bands under examination demonstrate the potential for sophisticated management of coverage and capacity trade-offs.

Note that the number of users is fixed at $10$ for this analysis, while in scenarios involving higher frequency bands, the potential for gMIMO to serve many more users becomes evident. This enhancement both improves the overall system performance and accentuates the advantages of operating within the upper mid-band frequencies, making them particularly attractive for advanced wireless communication systems. However, it is important to acknowledge that the observed higher DOF in the upper mid-band compared to sub-7 GHz bands may not solely be attributed to the frequency due to weaker multipath diversity itself but rather to the increased number of UE antennas under aperture size constraints. While this approach enhances spatial multiplexing, it also introduces power consumption and cost challenges, as antennas and transceiver chains become more expensive and energy-intensive at higher frequencies.

\section{NEAR-FIELD COMMUNICATIONS}
\label{S_Alva}

\begin{figure*}[t!]
\centering
\subfloat[One gMIMO array]{
\begin{overpic}[width=0.5\textwidth]{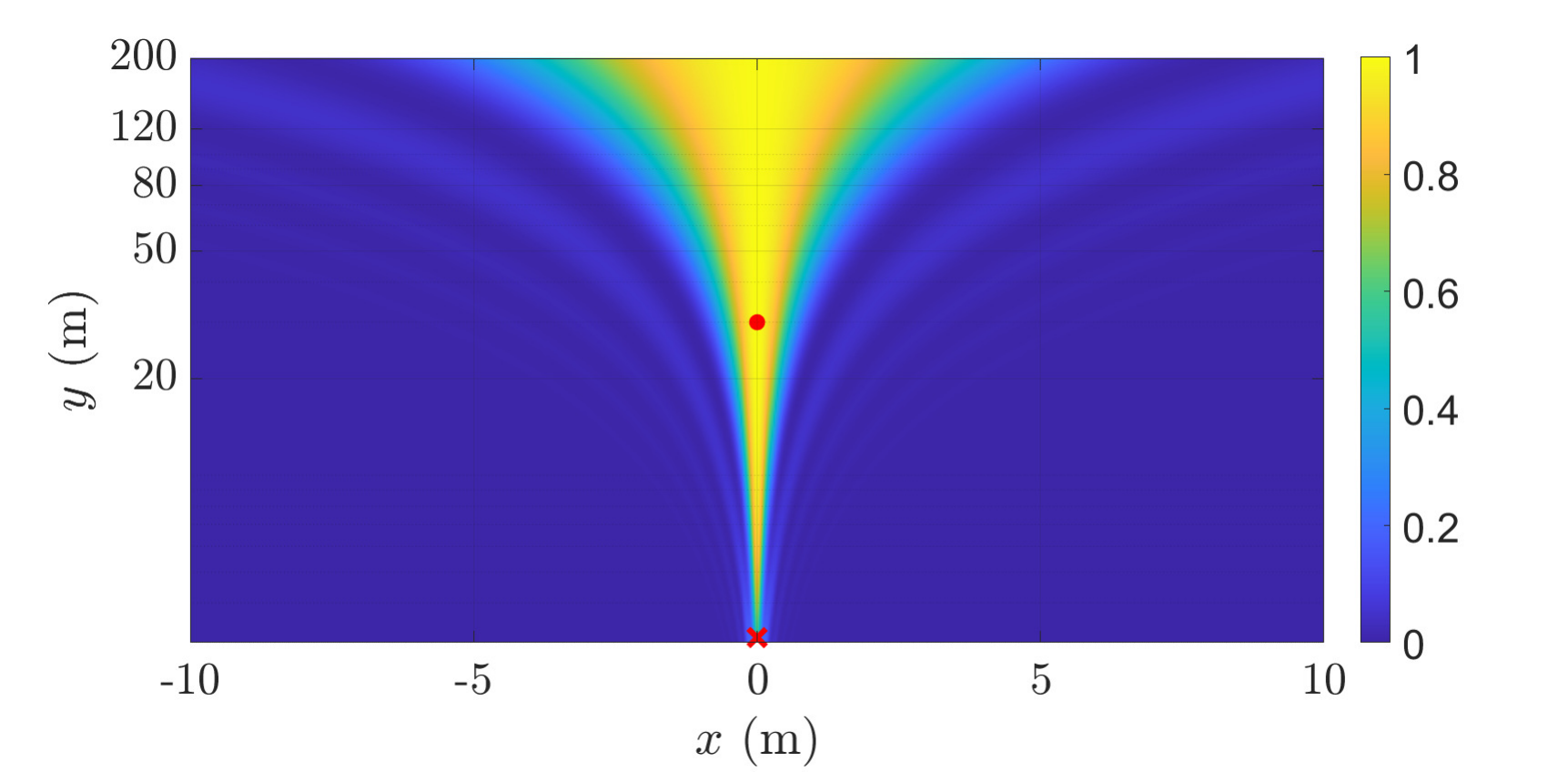}
\end{overpic}}
\subfloat[Two subarrays]
{\begin{overpic}[width=0.5\textwidth]{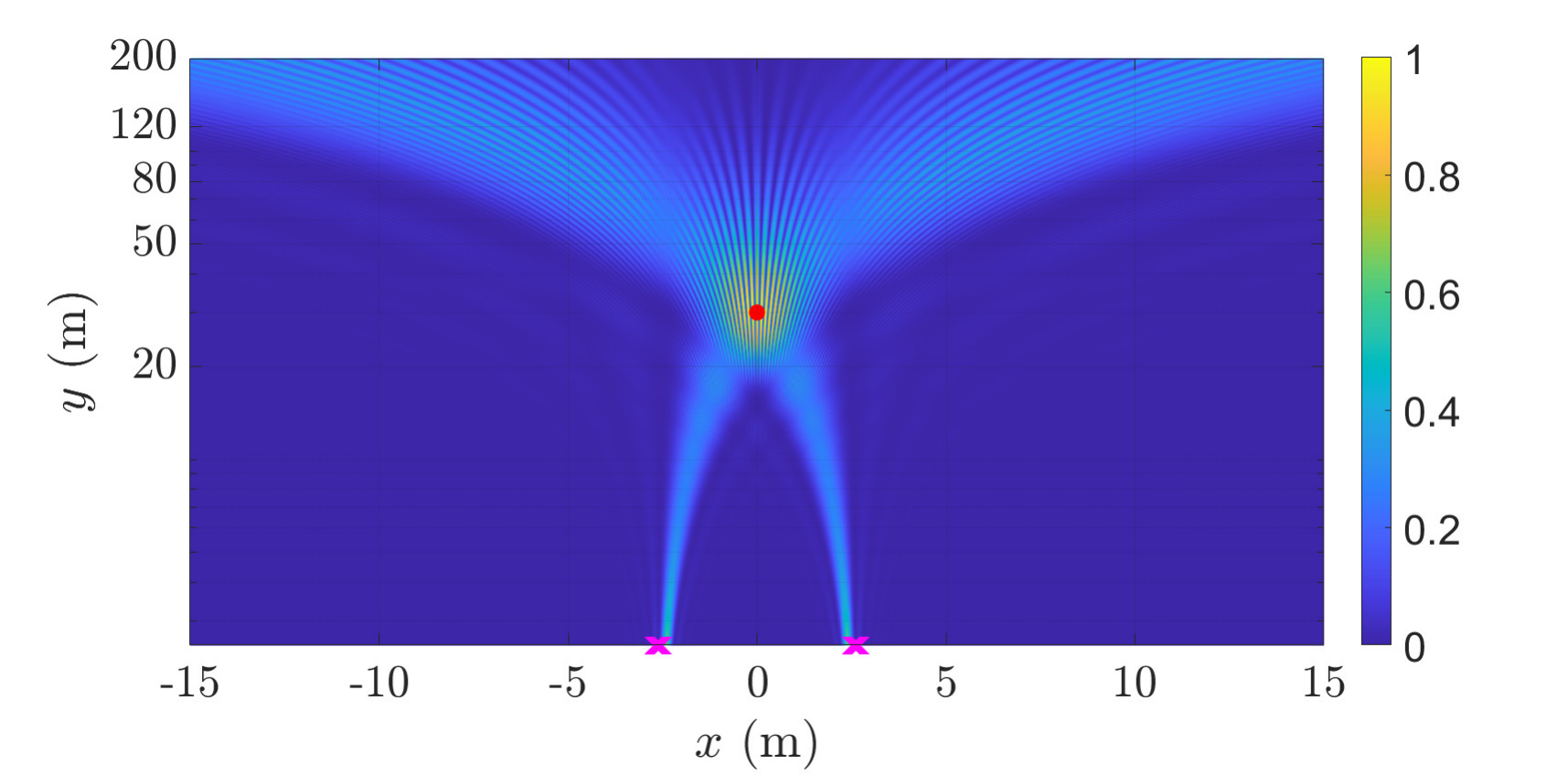}
\end{overpic}} \hfill
\subfloat[Illustration of one gMIMO array versus two subarrays deployed a few meters apart]{
\begin{overpic}[width=0.5\textwidth]{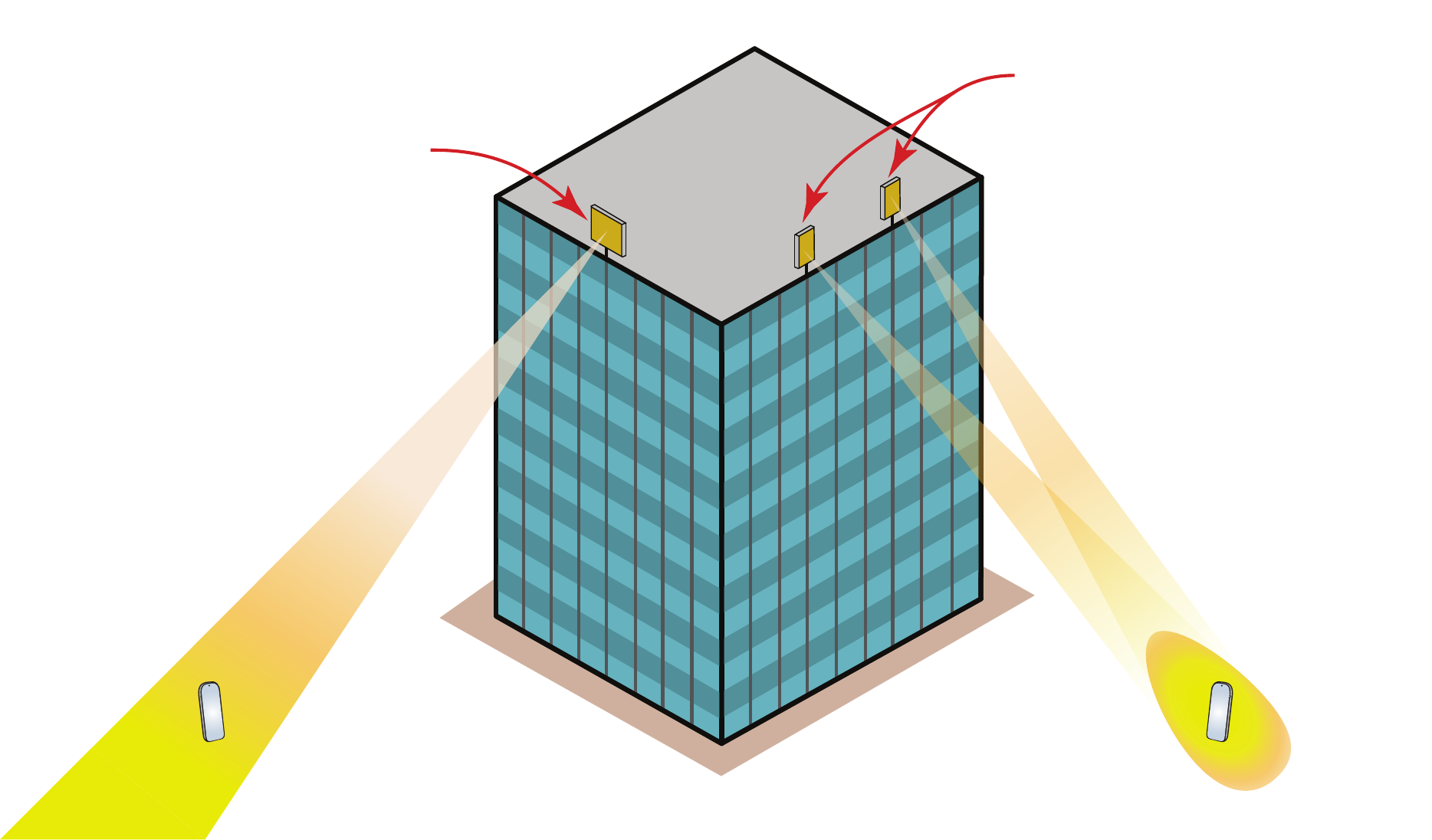}%
\put(13,52){\scriptsize One gMIMO array}%
\put(13,48){\scriptsize (48 antennas)}
\put(72,52.5){\scriptsize Two subarrays}%
\put(72,48.5){\scriptsize (24 antennas)}%
\end{overpic}}

\caption{Beampatterns in the upper mid-band depending on the array configuration: one array or two sub-arrays. The exact beampatterns are shown in (a) and (b), respectively, where the transmitter array locations are indicated with magenta ($\times$), while red (o) indicates the UE location (focal point). The colormap shows the normalized beamforming gain and is considered to be strong in the yellow/green regions. A sketch of the simulation setup and beampatterns is provided in (c).
}
\label{F_Bfc}
\end{figure*}

The beamwidth shrinks with the carrier frequency, as described in Section~\ref{S_Nikos}, but also the shape of the beam in the transverse propagation direction can change.
This phenomenon is associated with the so-called radiative near-field, which is the region where the transmitted wave's spherical curvature is noticeable at the receiver.
The far-field region with approximately planar wavefronts begins at the Fraunhofer distance, while the radiative near-field is the range between a few wavelengths from the transmitter (when reactive effects can be neglected) to the far-field.
The Fraunhofer distance is calculated as $ 2 D_{\rm array}^2/\lambda$, where $D_{\rm array}$ is the aperture length of the transmitter array \cite{Ramezani2023Exploiting}. 
When the aperture is physically large or the wavelength is tiny, the Fraunhofer distance can become large, and we will observe near-field propagation conditions when serving UEs at shorter distances. This is not the case in 5G, but we will explore if it occurs in 6G upper mid-band systems.

When a beam is focused on a near-field location, the transverse variations in spherical curvature dissolve the focusing after a particular distance.
Consequently, in the near-field, the beampattern resembles a spotlight, as opposed to the far-field beampattern that disperses as a cone across space \cite{Ramezani2023Exploiting}.
One significant advantage of such \emph{near-field beamfocusing} is that the BS can distinguish between user devices and multipath clusters in the distance domain, in addition to the angular domain. This feature makes multi-user MIMO channels appear richer and enhances the sum capacity. 

To explore the presence of such near-field effects in the prospective 6G bands, we plot the relative beamforming gain (from 0 to 1) observed in the azimuth plane in  Fig.~\ref{F_Bfc}(a).
The BS is equipped with a uniform linear array (ULA) composed of $N=48$ half-wavelength-spaced antennas operating at the $15$ GHz carrier frequency. This corresponds to the array in Fig.~\ref{fig:array_magazine_paper} but with one transceiver branch per column, since we simulate a two-dimensional setup.
The BS is at the origin and focuses a beam on a UE $30$\,m away in the broadside direction. The beamforming gain is large in the yellow and green regions, which represent the half-power beamwidth.\footnote{Note that we have used a logarithmic scale on the $y$-axis in Fig.~\ref{F_Bfc}(a) to show the beampattern over an extended distance interval. The angular beamwidth is the same at all distances, even if it looks like it expands due to the logarithmic scale.}
We observe that the beamforming gain continues to be strong behind the UE, and it will continue toward infinity. This is a typical far-field beampattern and was expected since the UE distance is larger than the Fraunhofer distance of $23$\,m in the considered setup.

To achieve near-field beamfocusing, we can increase the number of antennas in the array, thereby expanding the aperture length and Fraunhofer distance. Although this is theoretically possible, it is undesirable from a deployment and hardware complexity perspective.
Alternatively, we can shift the carrier frequency to the mmWave band, but that is unappealing due to poor coverage in non-LOS scenarios. Hence, we cannot use beamfocusing in 6G when considering conventional deployments and form factors.

There is one realistic way to exploit near-field beamfocusing in 6G: We can divide the array into smaller subarrays and deploy them at the same BS site but some meters apart. We can perform coordinated multi-array transmission and benefit from the expanded distance between the outermost antennas.
To demonstrate this possibility, we consider two subarrays with $24$ antennas, so the total number is $48$. The two arrays are separated by $5$\,m and focus the signal on the same UE location as before.
The corresponding beampattern is depicted in Fig.~\ref{F_Bfc}(b) and shows a precise beamfocusing effect at the desired location with minor energy leakage to the surrounding area.

These results suggest a new 6G deployment strategy for urban scenarios: instead of one conventionally sized BS panel, we can deploy multiple smaller panels separated by many meters on the same rooftop. They operate as a single BS and use a common baseband processing unit. It is a natural next step from current multi-sector BS sites, which also uses multiple panels but pointing in different directions.
The conceptual differences between the radiation patterns are shown in Fig.~\ref{F_Bfc}(c), where we sketch how the two subarrays transmit far-field-like beams that intersect at the UE and create the maximum beamforming gain only in a region around it. This phenomenon is reminiscent of how a human obtains depth perception using two eyes separated by many wavelengths.

Note that a very large array aperture can also be filled with a small number of antennas by deploying a ULA with a large antenna spacing, but this would lead to grating lobes \cite{massivemimobook}. By contrast, the proposed multi-panel deployment with half-wavelength-spaced antennas in each panel will only give rise to weak side-lobes.
The multi-panel deployment on a single BS site can be viewed as the first step towards the cell-free/distributed MIMO architecture \cite{massivemimobook}, where the arrays are distributed over large areas and require additional fronthaul and synchronization infrastructure.

\section{NEAR-FIELD LOCALIZATION AND SENSING}
\label{S_Parisa}

6G is envisioned to expand its functionality beyond connectivity and become an integrated communication, localization, and sensing service platform.
So far, we have demonstrated how gMIMO enables communication improvements in the upper mid-band spectrum, but can we also enable precise localization and environmental awareness from a single site? In this section, we will analyze these prospects. 

In the far-field, an antenna array can estimate the angle of a source transmitting a narrowband signal but not the distance since the planar wavefront lacks such information. In contrast, when the sources are within the radiative near-field region of the antenna array, high-resolution three-dimensional location estimation can be achieved by capturing both angle and distance information from the impinging spherical wavefronts \cite{Ramezani2025a}. 
These near-field features can be easily utilized in mmWave and sub-THz bands where the wavelength is tiny; however, the end of the near-field region (i.e.,  Fraunhofer distance) is only some tens of meters for a conventionally sized 6G BS in the upper mid-band.

In the Section~\ref{S_Alva}, we described how the deployment of multiple subarrays deployed with a few meters of separation enables near-field beamfocusing without requiring a huge continuous aperture. We will now demonstrate that localization can also benefit from this new deployment strategy. The individual subarrays observe locally planar wavefronts, but the spherical curvature is observable when fusing the received signals from multiple arrays.

\begin{figure}
\centering
\centering
\subfloat[Illustration of one gMIMO array versus four subarrays]
{
\begin{overpic}[width=0.5\textwidth]{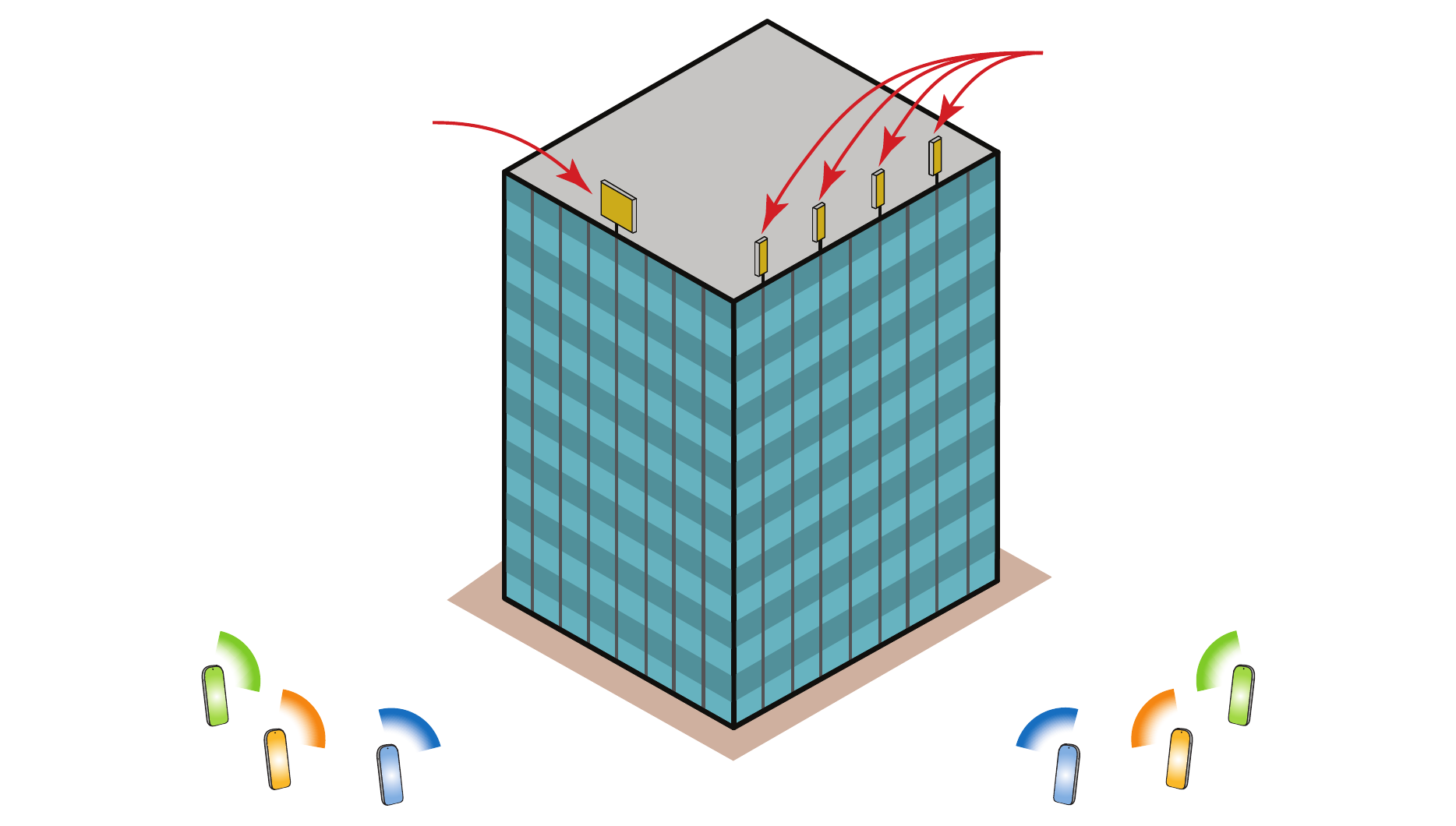}%
\put(13,52){\scriptsize One gMIMO array}%
\put(13,48){\scriptsize (48 antennas)}
\put(72.5,52.5){\scriptsize Four subarrays}%
\put(72.5,48.5){\scriptsize (12 antennas)}%
\end{overpic}}\hfill
\centering
\subfloat[One array with $48$ antennas]
{\includegraphics[width=\columnwidth]{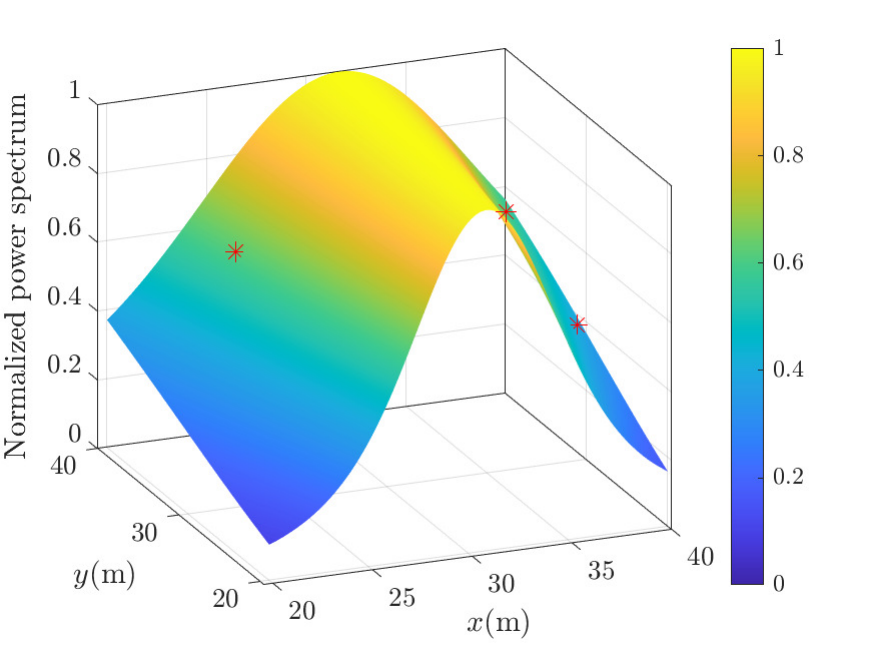}}\hfill
\centering
\subfloat[Four subarrays with $12$ antennas]
{\includegraphics[width=\columnwidth]{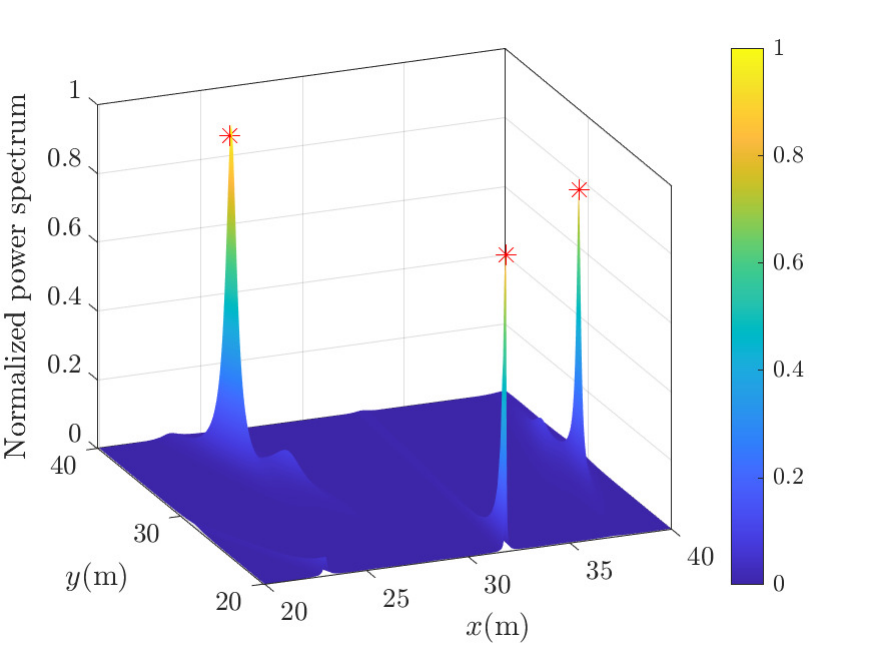}}
\caption{The MUSIC power spectra for localization of three closely spaced sources. A single conventionally-sized array at $15$ GHz cannot manage to resolve sources, but a subarray deployment can jointly estimate angles and distances.}
\label{fig:localization}
\end{figure}

Fig.\,\ref{fig:localization} considers the localization of three signal sources at the carrier frequency of $15\,\mathrm{GHz}$ using the well-known MUltiple SIgnal Classification (MUSIC) algorithm \cite{music}. We consider either a ULA with $48$ half-wavelength-spaced antennas\footnote{The result is also representative for a UPA with 48 antennas per row, so this is a gMIMO setup.} or the same number of antennas divided between four subarrays $5\,$m apart, deployed as shown in Fig.\,\ref{fig:localization}(a). 
Fig.\,\ref{fig:localization}(b) shows the normalized MUSIC power spectrum with one ULA. The MUSIC algorithm will fail to localize the sources since there are no peaks at their locations. This is expected since the sources are located beyond the Fraunhofer distance \cite{Ramezani2023Exploiting} and are closely spaced in angle. Therefore, the received signal looks like a single planar wave from which only one angle can be extracted.

Fig.\,\ref{fig:localization}(c) shows the corresponding power spectrum with four subarrays. We can see that the locations are accurately detected in this scenario thanks to the near-field propagation effects that the subarrays can jointly observe.
Specifically, the phase variations across the subarrays depend on each source's exact location due to the spherical wavefronts, making it possible to resolve the locations accurately. {This demonstrates that the gMIMO in the upper mid-band with a subarray structure can enable the sensing expectation of 6G as described in Section \ref{S_Ferdi} since cm-level resolution could be achieved.} 

In a nutshell, near-field localization using a single array necessitates an inconveniently large array in the upper mid-band due to the relatively large wavelength. Nevertheless, we can enable near-field localization using multiple small subarrays displaced by a few meters and processed jointly. These subarrays will receive the source signals from sufficiently different angles to localize them using triangulation-like techniques.
Similarly, for radar sensing, one subarray can form beams toward passive targets and the other subarrays can detect their locations in a multi-static fashion with high angle and distance resolution. One key difference from conventional localization and sensing systems that fuse information from multiple sources is that the subarrays are phase-synchronized.
This new deployment strategy is clearly suitable for integrated 6G near-field communication, localization, and sensing.

\section{OPEN RESEARCH AND STANDARDIZATION CHALLENGES}

Since the mid-band spectrum is limited, 6G must transition from mid-band to upper mid-band and from mMIMO to gMIMO to deliver vastly better functionalities than 5G. This has to be finalized in the short term, as the first 6G networks are expected to roll out by 2030. This section outlines the research challenges that must be overcome to enable this development, as well as the standardization efforts of the envisioned technology. This is the right time for the research community to seek answers to these scientific questions and affect the 6G standardization.

\subsection{Future Research Directions}
In the following, we identify seven research problems that need to be further studied to improve the performance of gMIMO systems. Considering the tight timelines of 6G standardization, enrichment in these areas may not be included in the first version but in later versions of 6G networks.

\textbf{1) New optimized antenna placement strategies:} 
We have demonstrated that communication and localization performance can be significantly enhanced by leveraging the near-field beamfocusing effect, which requires a sufficiently large array aperture when operating in the upper mid-band spectrum. This can be achieved through widely spaced subarrays, as previously shown, or alternatively futuristic movable antennas that offer dynamic configurations for specific tasks \cite{Zhu2024}. Both the subarray and the movable antenna approaches introduce irregularities in the array geometry, affecting algorithms that rely on the array response vector, which depends on both angle and distance in the near field. An antenna placement strategy, known as pre-optimized irregular array (PIA), was proposed in \cite{irshad2025} to achieve a performance comparable to that of movable antenna arrays. The PIA pre-optimizes the antenna locations in the array based on the intended coverage region and propagation environment around the BS to achieve better spatial resolution and diversity than conventional uniform planar array. Such deployments are more practically appealing than real-time movable antenna arrays that need to optimize the antenna locations in every channel realization and have extra hardware components for mechanical/electrical antenna movement.
However, these preliminary results point toward many unanswered questions: What is the qualitative interplay between the desired array geometry and the user distribution? How to determine the best antenna array geometry for a given propagation scenario? What is the role of near-field and far-field effects on the antenna placement? How can out-of-band interference be controlled when using irregular arrays? How do the frequency characteristics (i.e., upper mid-band) affect the answers?

\textbf{2) Medium-resolution transceivers and computations:} We need more antennas to maintain FR1-like propagation conditions in the upper mid-band, and many transceiver branches to control them. Since hybrid analog/digital transceivers are unsuitable for multi-user scenarios and one-bit digital transceivers cannot coexist well with other systems, we must develop a new generation of medium-resolution MIMO transceivers. Lowering the hardware resolution by just a few notches can drastically reduce cost and energy consumption at the expense of non-linear phase injections, amplification, and calibration errors. This calls for new algorithms, possibly empowered by ML, to track hardware characteristics. Moreover, the multi-user beamforming complexity grows cubically with a fixed antenna/user ratio \cite{massivemimobook}, which might be managed through novel digital compression and offloading methods. Metamaterials might also play a role in future transceivers.
Furthermore, to align with the O-RAN (a strategic plan widely accepted by both academia and industry), it is essential to explore the feasibility of developing a quantized version of beamforming at the distributed unit (expected to be low-cost), with the central unit managing all complex signal processing tasks.

\textbf{3) Resource-efficient channel estimation:} Beamforming and spatial multiplexing require a timely and precise CSI, most efficiently acquired by sending one distinct pilot sequence per spatial DOF per channel coherence block \cite{massivemimobook}. We need longer pilots to accommodate more DOF in 6G, but the coherence time generally shrinks with wavelength. In other words, we need more pilot resources than in 5G, but start with having less.
New resource-efficient estimations are needed to overcome this challenge, possibly through pilot reuse within each cell and algorithmic addressing of the resulting contamination. A related challenge is to boost the channel estimation quality per DOF because the smaller antenna sizes lead to a reduced SNR per antenna in the upper mid-band. As described in Section~\ref{S_Nikos}, we can effectively counteract this effect through beamforming, but we can only exploit those gains once CSI has been acquired and not in the estimation phase. Parametric estimation/localization/sensing methods might be utilized to this end, but it remains to be shown whether these methods can be enhanced to function in practical complex environments and not only idealized LOS scenarios.

\textbf{4) Information-theoretic capacity limits:} 
The theoretical MIMO foundations are deep for single-antenna UEs \cite{massivemimobook}, but there are many open fundamental problems for scenarios with multi-antenna UEs. What are the DOF of a noncoherent MIMO system with structured channels? The seminal work \cite{Zheng2002a} only covers isotropic fading, which is a pessimistic assumption, yet we follow its guidelines when designing current networks. Moreover, the papers \cite{marzetta,BHS18A} from the mMIMO research era give two different perspectives on the role of pilot contamination on the fundamental capacity limits: it is the main limiting factor in isotropic fading, while it can be overcome using spatial channel correlation knowledge in other scenarios.
Some of the pertinent open research questions are:
What information-theoretic DOF can be achieved in a practical non-isotropic fading environment, and how does the answer depend on the available statistical information?
What kind of waveform and combination of data/reference signals are sufficient to fully exploit the available DOF? Does the choice of multi-carrier modulation format have an impact on these fundamental limits?
Is there a need for non-linear processing or rate-splitting methods?

\textbf{5) Energy-efficiency optimization:}
gMIMO has promising implications for enhancing energy efficiency by compensating for the increased energy consumption of having additional antenna hardware and increased signal processing complexity with significantly higher data rates. However, more detailed studies are needed to fully understand these gains. 
Dynamic sleep mode strategies can further reduce energy consumption by activating/deactivating antennas based on user demand.
Furthermore, while multi-panel transmission and localization improve coverage and accuracy without adding more antennas (recall Sections~\ref{S_Alva} and \ref{S_Parisa}), the synchronization between sub-arrays may increase energy consumption, making it essential to closely analyze the energy efficiency from many viewpoints. It is essential to take a holistic view on energy-efficiency optimization, including the deployment architecture \cite{Bjornson2015a}, the hardware components \cite{Rappaport2024a}.

\textbf{6) Integration of sensing and communication infrastructures:} In 6G networks, communication, and sensing operations are expected to be seamlessly integrated. Designing this integration smoothly and efficiently is of great importance. However, conventional radar systems typically operate at high carrier frequencies, whereas the potential 6G carrier frequency in the upper mid-band may not provide the necessary resolution for radar systems. Therefore, innovative waveform designs and receiver processing algorithms that can simultaneously meet the performance requirements for both sensing and high communication rates are needed. This presents an open research challenge in the development of 6G networks, where achieving a balance between high-resolution sensing and high-speed communication remains a critical objective. 

Furthermore, integrating information from various sensors, such as cameras, inertial measurement units, and environmental sensors, with communication signals can significantly improve the accuracy and robustness of localization. These sensors provide complementary information, enabling high-resolution positioning even in complex environments where traditional sensing methods at upper mid-band frequencies may be inadequate. By leveraging this approach, 6G networks can achieve a seamless balance between high-throughput communication and accurate sensing.

\textbf{7) Interplays with Other Emerging Technologies:}
gMIMO in the upper mid-band will certainly be the central component in the 6G radio access networks, and we demonstrated that ISAC and near-field communications could be integral parts of it. However, the research community has also put a lot of effort into other emerging technologies, such as reconfigurable intelligent surfaces (RISs), integration between terrestrial and non-terrestrial networks, orbital angular momentum, quantum communication, etc.  Is there any synergy between gMIMO and new emerging technologies? If so, how can they be exploited under practical circumstances? For instance, the authors in \cite{upper_mid_band_RIS} analyzed possible RIS deployments in the 6G upper mid-band that can support specific MNO challenges. If any of these emerging technologies should play a major role in 6G, now is the right time to start integrating them into the core 6G technology; that is, upper-mid band gMIMO functionalities.

\subsection{6G Standardization and Practical Implementation} 
We expect that 3GPP will focus its 6G standardization efforts on the lower and upper mid-band, as many companies envision the 7--24 GHz range as the 6G spectrum. This is supported by the technical documents from the 3GPP 6G Workshop in March 2025 \cite{ATT_6GWS,Apple_6GWS,Nokia_6GWS,Ericsson_6GWS}. Moving to the new frequency range requires significant modifications from the existing standards (legacy). 
We will give key examples below.

The UE antennas are assumed to be isotropic and have half-wavelength spacing in the legacy. However, this assumption does not hold in practice, particularly not in the upper midband, where the combined antenna radiation pattern becomes more complex and chaotic. As a result, additional adjustments to signal processing will be necessary, such as re-inventing the codebook design and making it more flexible to handle highly irregular array deployments in both the BSs and UEs.

Another important aspect of operating at 7--24 GHz is managing a large number of antenna ports. Increasing the channel state information reference signal (CSI-RS) to support 128 antenna ports in Release 19 was an effort driven by improving channel estimation, beamforming accuracy, and scalability for larger arrays with a large number of antennas. This trend is expected to continue towards gMIMO in Release 20, which will be the first 6G release targeting 256 or even 512 ports of CSI-RS.  However, handling such a large number of ports introduces significant complexity and overhead in CSI reporting. 
The various existing 5G methods (e.g., Codebook Type I and Type II, along with all their variations) require consolidation to streamline options in 6G standardization. The uplink channel estimation features using the sounding reference signals (SRS) must also be revisited to support spatial multiplexing of more UEs, while boosting the SNR in the channel estimation algorithms. This is related to the previously mentioned open research challenges, but there are also features from the prior mMIMO literature that remain to be standardized and implemented.

Furthermore, modifications to the legacy channel models are necessary when considering large arrays. 
This includes capturing radiative near-field effects, where the channel accounts for spherical wavefront curvature in both line-of-sight (LoS) 
and non-LoS (NLoS) conditions, as well as non-stationary effects due to variations in propagation across different parts of the array. Furthermore, the models must incorporate more refined delay and Doppler effects, as Doppler shifts in near-field propagation can vary across the antennas in the array. 
To date, there are no widely accepted near-field channel models. Addressing these challenges will require extensive measurements, enhanced ray-tracing techniques, and machine learning to develop minimally complex models that accurately capture spherical wavefront propagation and environmental interactions. These models will be used to refine MIMO signal processing algorithms to exploit these features effectively.

\section{CONCLUSIONS}

The first 6G networks will operate in the upper mid-band with modest spectrum availability but great prospects for huge-dimensional beamforming and spatial multiplexing, both in theory and practice. We call this paradigm \emph{Gigantic MIMO} and have shown how it could be used to reach the 6G performance goals. We have demonstrated how distributed subarray deployment enables near-field beamfocusing and localization/sensing, which would otherwise not be possible in the upper mid-band. Finally, pertinent research, standardization, and deployment challenges have been identified.

\bibliographystyle{IEEEtran}
\bibliography{refs}

\begin{IEEEbiography}
[{\includegraphics[width=1in,height=1.25in,clip,keepaspectratio]
{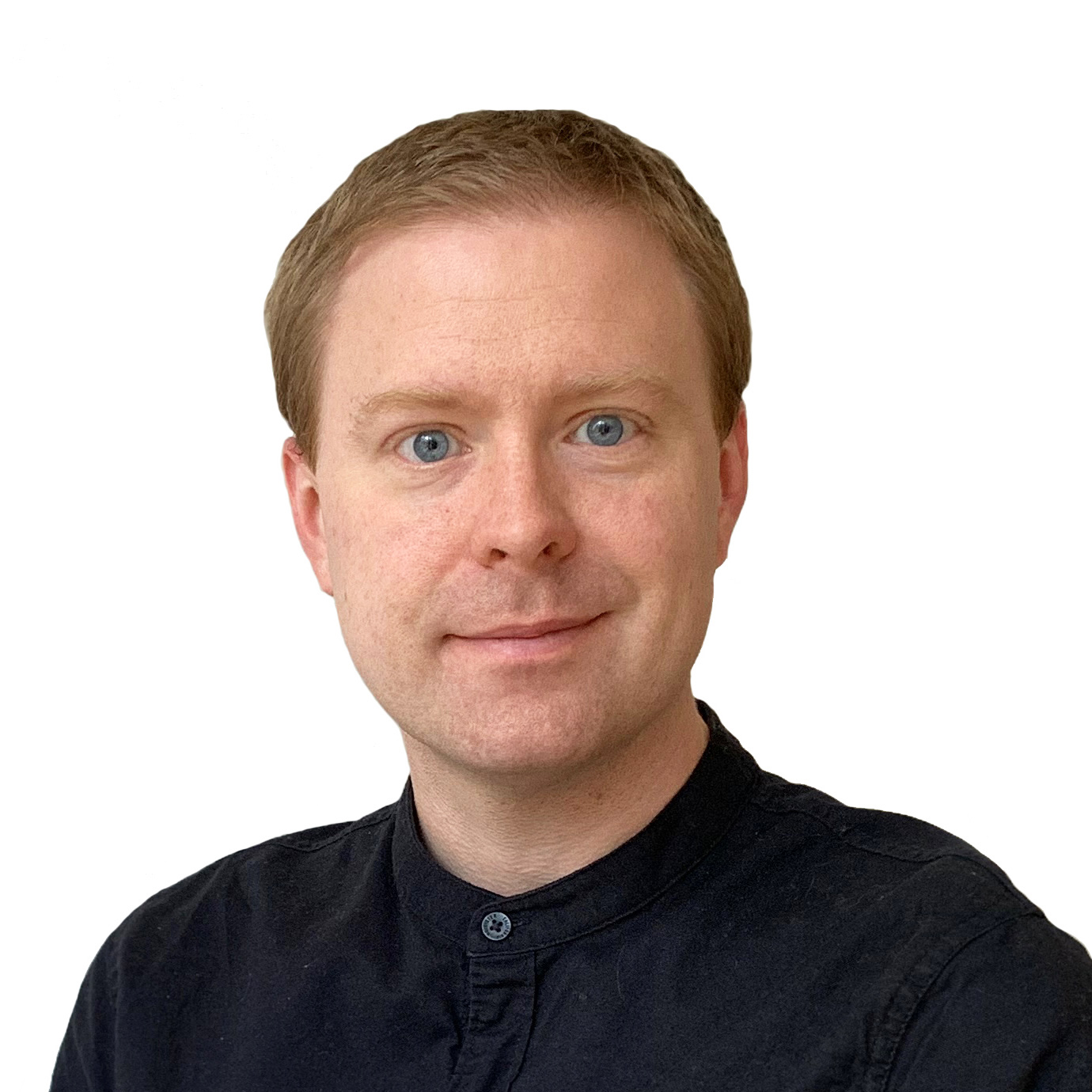}}]{Emil Bj\"ornson }
received the M.S. degree in engineering mathematics from Lund University, Sweden, in 2007, and the Ph.D. degree in telecommunications from the KTH Royal Institute of Technology, Sweden, in 2011. He is a Professor of Wireless Communication at the KTH Royal Institute of Technology, Stockholm, Sweden. He is an IEEE Fellow, Digital Futures Fellow, Wallenberg Academy Fellow, and Clarivate Highly Cited Researcher. His research focuses on multi-antenna communications and radio resource management, using methods from communication theory, signal processing, and machine learning. He has a podcast and YouTube channel called Wireless Future, has authored four textbooks, and published much simulation code. 
\end{IEEEbiography}

\begin{IEEEbiography}
[{\includegraphics[width=1in,height=1.25in,clip,keepaspectratio]
{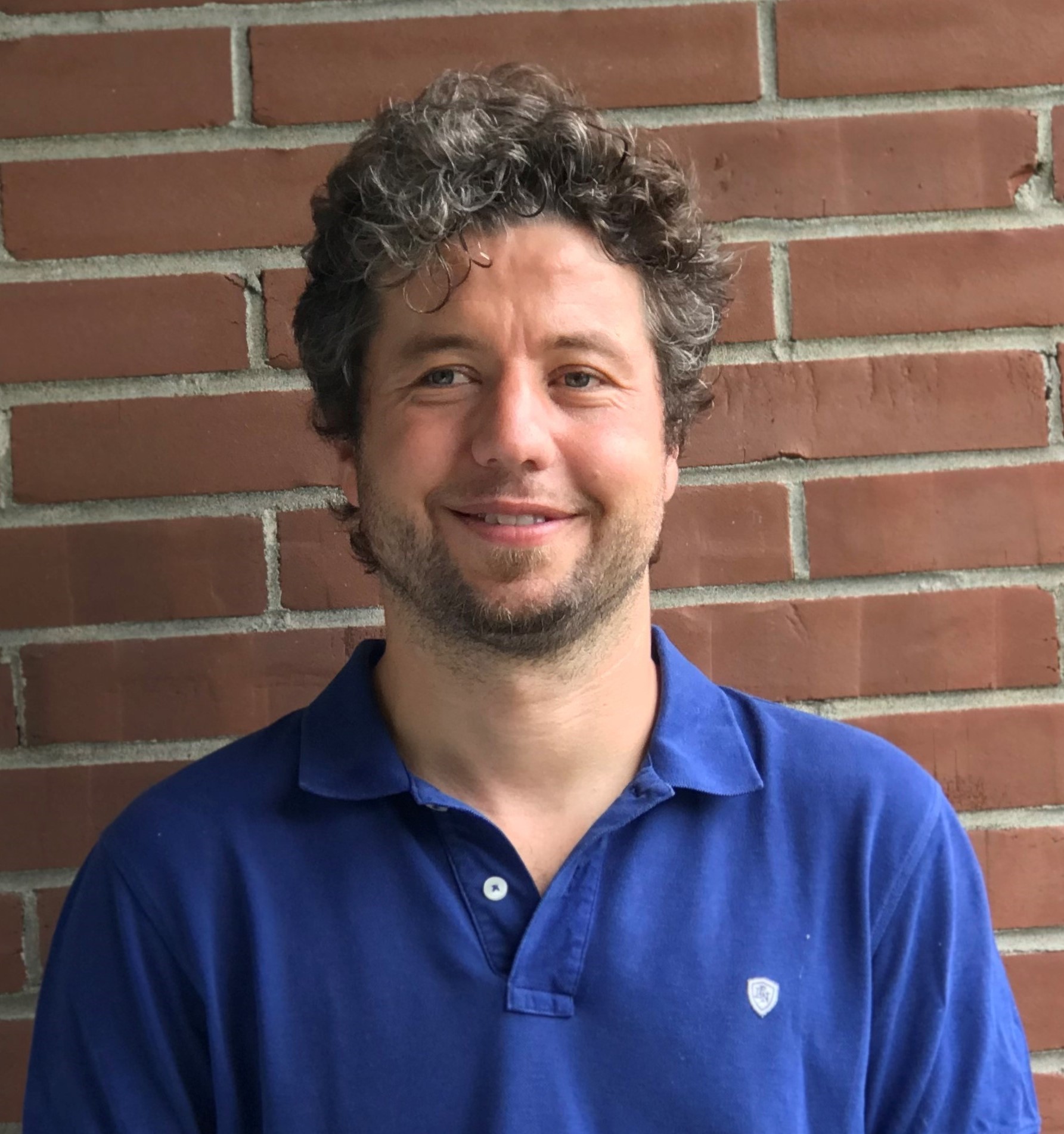}}]{Ferdi Kara } received Ph.D. degree in Electrical and Electronics Engineering from Zonguldak Bulent Ecevit University (ZBEU), Turkiye, in 2019. Since 2011, he held several positions at the  ZBEU and is now an Associate Professor at Computer Engineering and also a postdoctoral researcher at the KTH Royal Institute of Technology, Stockholm, Sweden. From 2021 to 2023, he was a Senior Research Associate at Carleton University, Ottawa, Canada. He is the recipient of several prestigious awards and honors, including the 2024 Outstanding Young Researcher Award of the IEEE Communication Society Europe, Middle East, and Africa (EMEA) Region. He is an editor for IEEE Open Journal of Communications Society, IEEE Communications Letters, EURASIP Journal of Wireless Communications and Networking, and Physical Communication (Elsevier). His research interests span wireless communications and networking, with a particular focus on physical layer (PHY) aspects. 

\end{IEEEbiography}

\begin{IEEEbiography}
[{\includegraphics[width=1in,height=1.25in,clip,keepaspectratio]
{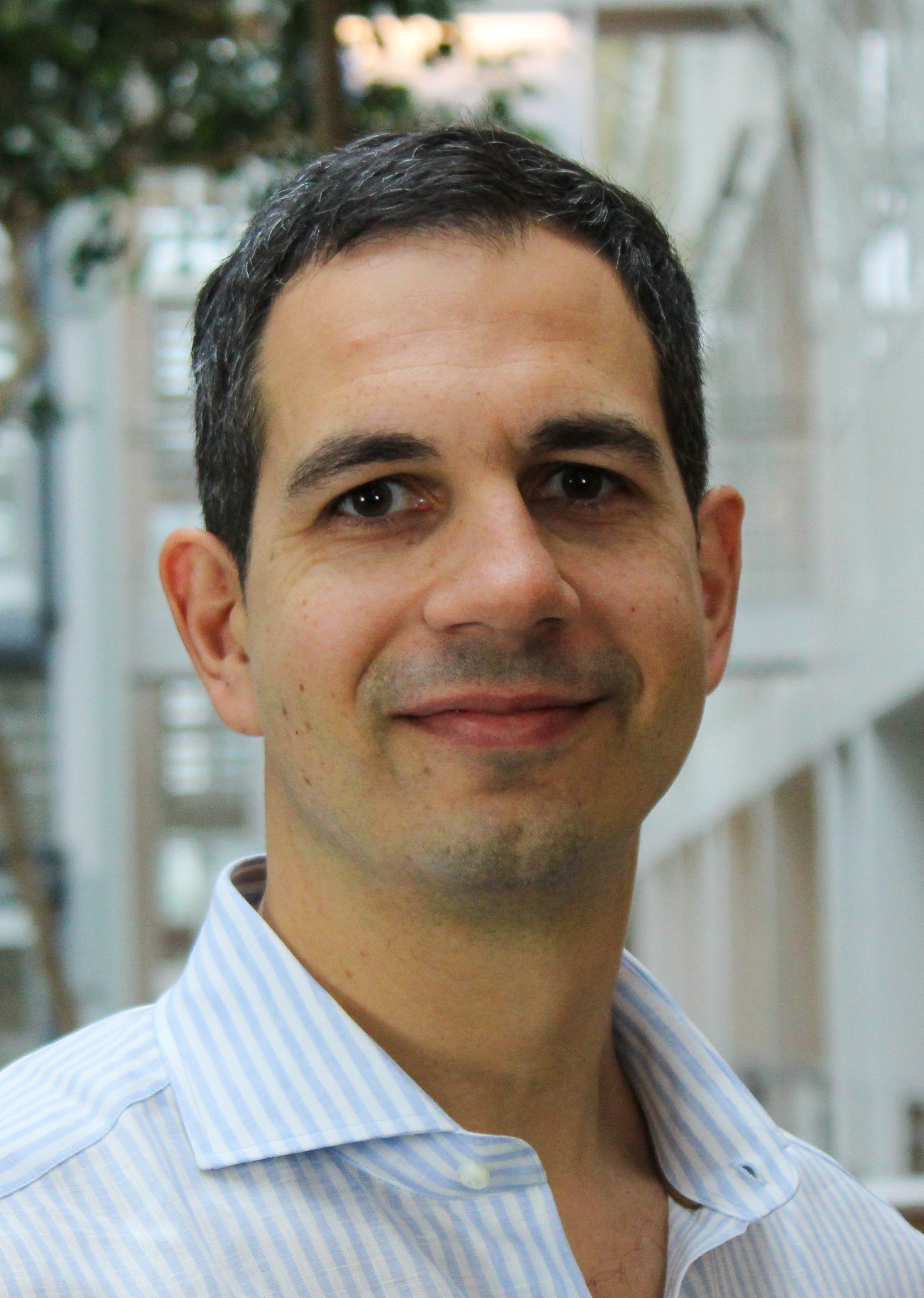}}]{Nikolaos Kolomvakis }
received the M.Sc. degree in Electrical Engineering and Information Technology from the Swiss Federal Institute of Technology (ETH) Zurich, Switzerland, in 2012, and the Ph.D. degree in Wireless Communications from Chalmers University of Technology, Gothenburg, Sweden, in 2019. From 2017 to 2023, he was a Systems Engineer and Senior Researcher at Ericsson AB in Stockholm, Sweden. Since 2023, he has been a Researcher in the Division of Communication Systems at KTH Royal Institute of Technology, Stockholm, Sweden and a Visiting Researcher at Ericsson AB, Stockholm, Sweden. His research interests include wireless communications and signal processing, with a particular emphasis on multi-antenna technologies.
\end{IEEEbiography}

\begin{IEEEbiography}
[{\includegraphics[width=1in,height=1.25in,clip,keepaspectratio]
{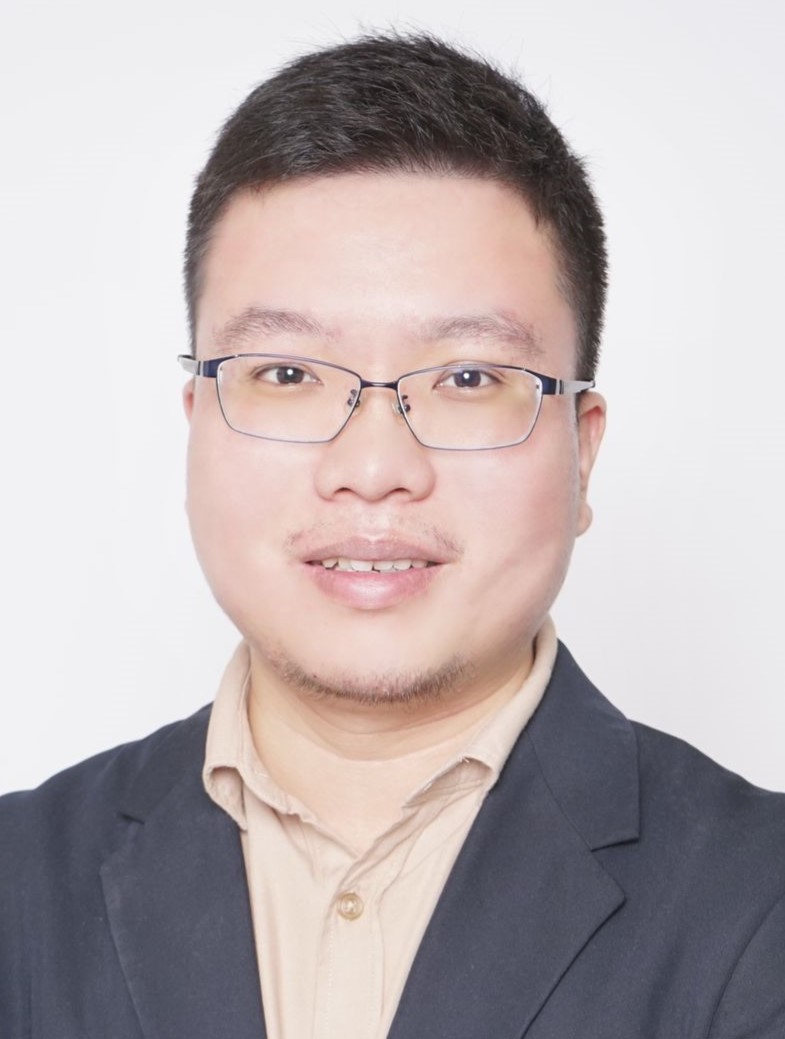}}]{Alva Kosasih }
received his B.Eng. and M.Eng. degrees in Electrical Engineering from Brawijaya University, Indonesia, in 2013 and 2017, respectively, his M.S. degree in Communication Engineering from National Sun Yat-sen University, Taiwan, in 2017, and his Ph.D. in Communication Engineering from the University of Sydney, Australia, in 2023. From 2023 to 2024, he was a Postdoctoral Researcher at KTH Royal Institute of Technology, Stockholm, Sweden. He is currently a 3GPP RAN1 delegate at Nokia Standards. His research interests include signal processing for large-scale MIMO; near-field beamfocusing, localization, and channel estimation; MIMO symbol detection; and machine learning for physical-layer communications.
\end{IEEEbiography}

\begin{IEEEbiography}[{\includegraphics[width=1in,height=1.25in,clip,keepaspectratio]{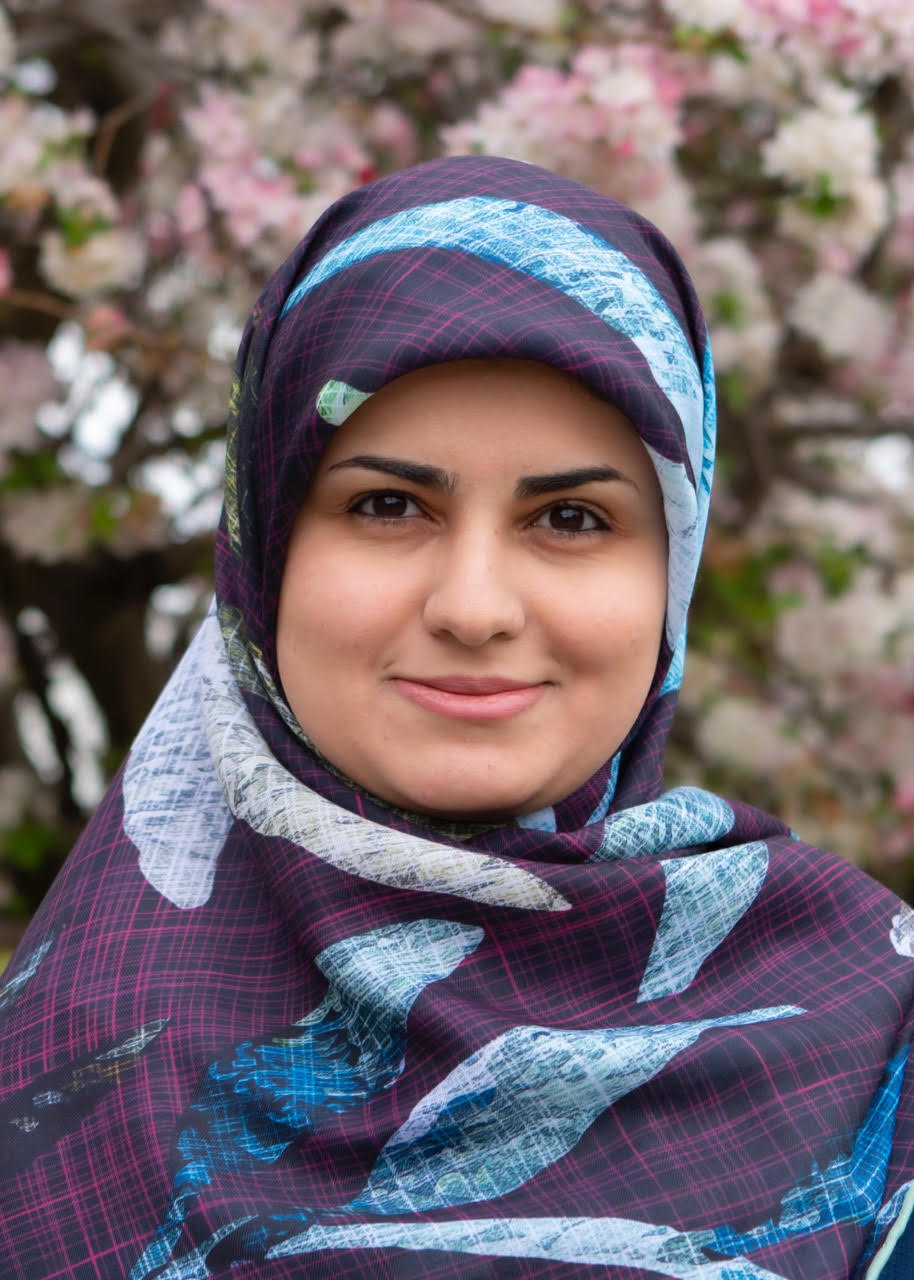}}]{Parisa Ramezani } 
received her B.Sc. degree in electrical engineering from Sharif University of Technology, Tehran, Iran, in 2014, and her M.Phil. and Ph.D. degrees in electrical and information engineering from the University of Sydney, Sydney, Australia in 2017 and 2022, respectively. Since 2022, she has been a postdoctoral researcher at KTH Royal Institute of Technology, Stockholm, Sweden. She has been recognized as an Exemplary Reviewer for the IEEE Transactions on Communications in 2022. Her research interests lie in the fields of wireless communications and signal processing, with a focus on reconfigurable intelligent surface assisted communications, near-field localization and beamforming, and backscatter communications.
\end{IEEEbiography}

\begin{IEEEbiography}[{\includegraphics[width=1in,height=1.25in,clip,keepaspectratio]{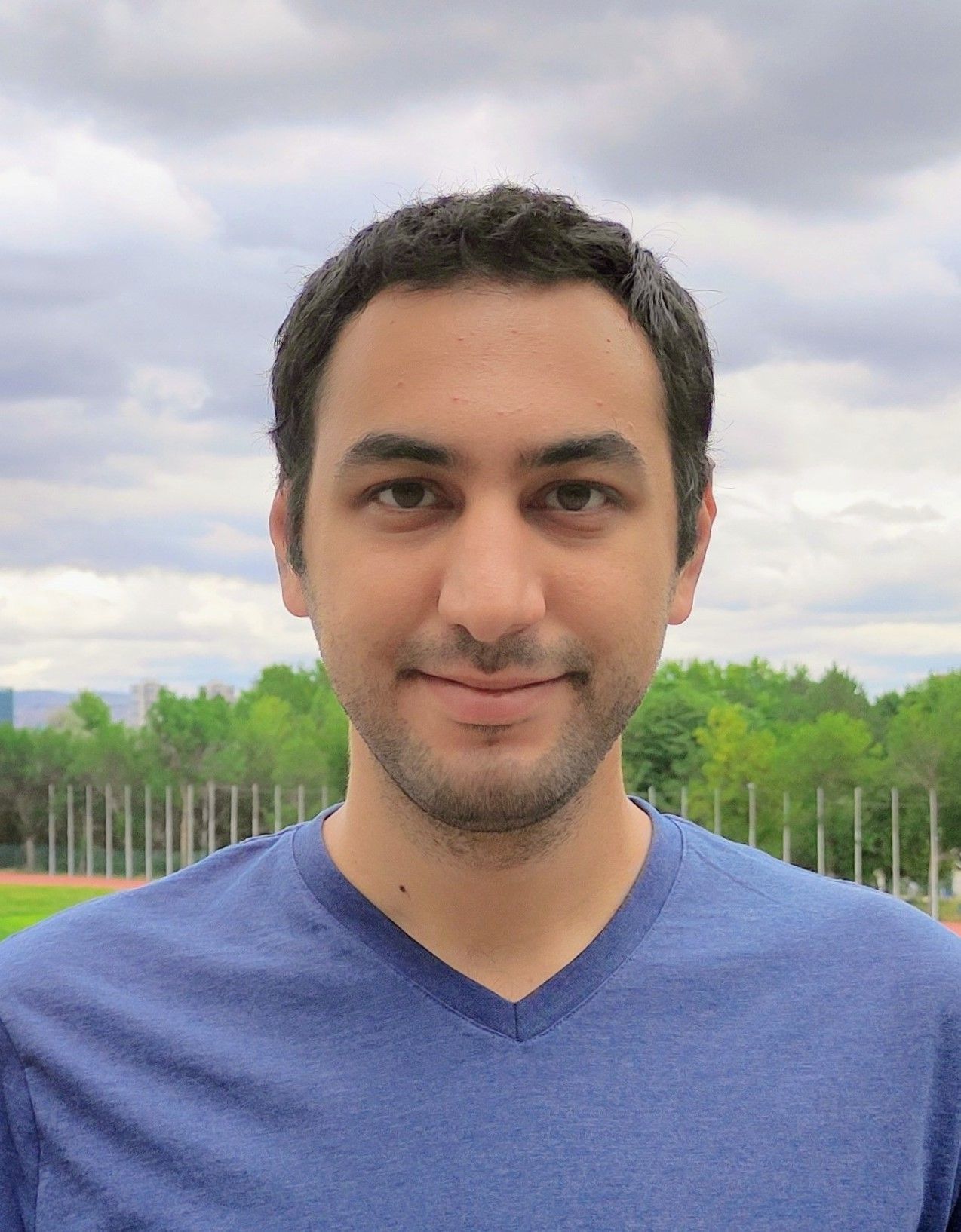}}]{Murat Babek Salman } 
received B.S., M.S., and Ph.D. degrees both in electrical and electronics engineering from the Middle East Technical University, Ankara, Turkey, in 2015, 2018 and 2023, respectively. Since 2023, he has been with the Communications Systems Division, KTH Royal Institute of Technology as a post-doctoral researcher. From 2022 to 2023, he held a Visiting Researcher position at KTH Royal Institute of Technology under the Department of Communication Systems. His research interests include adaptive signal processing, signal processing for wireless communications with a particular focus on transceiver nonlinearities, full duplex communications, integrated sensing, and communications.
\end{IEEEbiography}

\end{document}